\documentclass[preprint]{revtex4-1}

\usepackage{graphicx}
\usepackage{bm}
\usepackage{scalerel}
\usepackage[utf8]{inputenc}
\usepackage[T1]{fontenc}
\usepackage{mathptmx}
\usepackage{xcolor}
\usepackage{amsmath}
\usepackage[colorlinks=true, citecolor=blue,
  linkcolor=blue]{hyperref}
\newcommand{\Rey}{Re}
\begin{document}

\preprint{AIP/123-QED}

\title[Hydrodynamic interactions between sphere and wall]{Dynamics of an internally actuated  weakly elastic sphere translating parallel to a rigid wall\\}

\author{Shashikant Verma}
\affiliation{%
Department of Mechanical Engineering, Indian Institute of Technology, Ropar 140001, India
}%

\author{B. Dinesh}
\affiliation{%
Department of Chemical Engineering and Technology, Indian Institute of Technology (BHU), Varanasi 221005, India
}%

\author{Navaneeth K. Marath}
\affiliation{%
Department of Mechanical Engineering, Indian Institute of Technology, Ropar 140001, India
}%

\date{\today}

\begin{abstract}
We analyse the dynamics of a weakly elastic spherical particle translating parallel to a rigid wall in a quiescent Newtonian fluid in the Stokes limit. The particle motion is constrained parallel to the wall by applying a point force and a point torque at the centre of its undeformed shape. The particle is modelled using the Navier elasticity equations. The series solutions to the Navier and the Stokes equations are utilised to obtain the displacement and velocity fields in the solid and fluid, respectively. The point force and the point torque are calculated as series in small parameters $\alpha$ and $1/H$, using the domain perturbation method and the method of reflections. Here, $\alpha$ is the measure of elastic strain induced in the particle resulting from the fluid's viscous stress, and $H$ is the non-dimensional gap width, defined as the ratio of the distance of the particle centre from the wall to its radius. The results are presented up to $\textit{O}(1/H^3)$ and $\textit{O}(1/H^2)$, assuming $\alpha \sim 1/H$, for cases where gravity is aligned and non-aligned with the particle velocity, respectively. The deformed shape of the particle is determined by the force distribution acting on it. The hydrodynamic lift due to elastic effects (acting away from the wall)  appears at $\textit{O}(\alpha/H^2)$, in the former case. In an unbounded domain, the elastic effects in the latter case generate a hydrodynamic torque at \textit{O}($\alpha$) and a drag at \textit{O}($\alpha^2$). Conversely, in the former case, the torque is zero, while the drag still appears at \textit{O}($\alpha^2$).
\end{abstract}

\maketitle


\section{Introduction}\label{sec:introduction}
\label{sec:headings}

Quantifying the hydrodynamic force and torque acting on a particle moving in confined flows is crucial for various industrial and biological applications such as the separation of cells in microfluidic devices \citep{secomb1986flow,wu2021film,song2023recent}.  In many of these applications, the ambient flow is predominantly unidirectional. In such flows, it is well known that a rigid spherical particle does not experience lateral movement in the Stokes limit due to the Stokes reversibility \citep{guazzelli2011physical}. Many analytical studies have examined the dynamics of the particle near a rigid wall and found no lateral movement in the Stokes regime \citep{dean1963slow,o1964slow,goldman1967slow}. However, the particle can move laterally due to non-linear effects like inertia (finite Reynolds number flows) and fluid non-Newtonian properties \citep{matas2004lateral}. In shear flow, the inertial effects can induce lateral migration of the particle through three distinct contributions: 1) due to the slip velocity of the particle \citep{saffman1965lift}, 2) due to the interactions with the wall, and 3) due to the gradient in the ambient shear flow \citep{ho1974inertial,vasseur1976lateral,anand2023inertial}. These contributions explain several experimental observations, including the Segre-Siberberg effect \citep{segre1961radial,matas2004lateral,nakayama2019three}. In contrast, a deformable particle, such as a spherical drop/bubble in wall-bounded flows, exhibits lateral migration even in the Stokes regime due to deformation-induced wall lift force \citep{leal1980particle}. In a single wall-bounded linear shear flow, a neutrally buoyant spherical drop exhibits migration away from the wall at a rate varying inversely to the square of the distance between the wall and the drop \citep{chaffey1965particle}. In general, a drop of arbitrary density and viscosity experiences an inertial-induced and deformation-induced wall lift force while translating next to a rigid wall \citep{magnaudet2003drag}. The dynamics of a rigid particle moving next to a flexible surface have been analysed using analytical, numerical and experimental methods \citep{urzay2007elastohydrodynamic,saintyves2016self,rallabandi2018membrane,bertin2022soft,rallabandi2024fluid}. The lift force on a rigid cylinder and sphere moving near an elastic wall has been estimated in the Stokes limit \citep{rallabandi2017rotation,daddi2018reciprocal}. Elastohydrodynamic interactions between soft material-coated interfaces have been studied within the lubrication approximation \citep{sekimoto1993mechanism,skotheim2005soft}. The lift force has been evaluated for various geometrical configurations, including: a rigid cylinder moving parallel to a soft-coated rigid substrate, a soft cylinder moving parallel to a rigid substrate, a cylindrical shell moving parallel to a rigid substrate, and a cylindrical journal bearing coated with a soft layer \citep{skotheim2005soft}. In all the configurations, the soft material has been modelled as a linear elastic, compressible solid governed by Navier elasticity equations. Further, the influence of substrate viscoelasticity on the lift force experienced by a rigid sphere translating parallel to a soft surface has been studied analytically in the lubrication limit \citep{kargar2021lift}.
  
 In the Stokes limit, the lateral movement/lift force observed when either the wall or the particle is deformable arises due to a breakdown of the Stokes reversibility. The effects of particle deformability have been extensively studied through numerical simulations, particularly for vesicles, capsules, and drops \citep{doddi2009three,nix2014lateral,barthes2016motion,bureau2023lift}. 
However, modelling the particle as an elastic solid rather than a drop alters its dynamics due to differences in their constitutive equations. Recent experiments have demonstrated that wall effects suppress the terminal velocity of an elastic particle, whereas its deformability enhances it \citep{noichl2022dynamics}. To model elasticity, one can use Hooke's law for small deformations and the Neo-Hookean model \citep{li2020understanding} for finite deformations. Theoretical studies have showed that the dynamic behaviour of the particle is sensitive to the chosen model; for instance, in an unbounded domain, a compressible Hookean sphere exhibits enhanced terminal velocity  \citep{murata1980deformation}, whereas an incompressible Neo-Hookean sphere experiences a reduction in terminal velocity compared to the corresponding Stokes velocity \citep{nasouri2017elastic}. The dynamics of rigid spheres, vesicles, capsules, and drops near walls are well-studied, but analytical investigations of wall effects on elastic solids remain limited. Most analytical studies have focused on elastic solids moving in an unbounded domain \citep{tam1973transverse,murata1980deformation}, leaving a knowledge gap in wall interaction effects. For a rigid sphere moving near a rigid wall, analytical techniques using bispherical harmonics \citep{o1964slow}, lubrication theory \citep{goldman1967slow}, and the method of reflections\citep{kim2013microhydrodynamics} provide solutions in the Stokes limit. The results from the method of reflections for the sphere align well with experimental data for wall-particle gap widths greater than 1.5 (translation) and 1.3 (rotation) \citep{malysa1986rotational}. Further, it has been observed numerically that at low capillary number, defined as the ratio of viscous and elastic forces, the lift velocity of a capsule in a bounded simple shear flow is nearly equal to the analytical solution (based on far-field approximation) over a wide range of wall–particle gap widths \citep{nix2014lateral}.
\begin{figure}
\centerline{\includegraphics[width=0.73\textwidth]{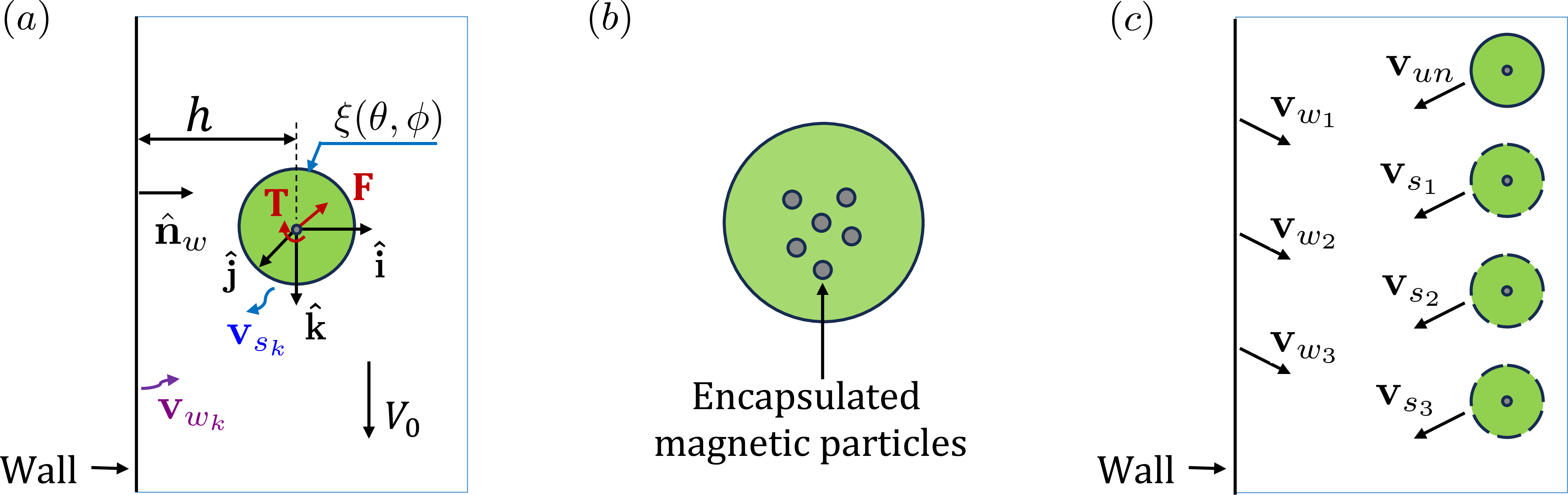}}
  \caption{(a) Schematic of an elastic sphere translating parallel to a rigid wall at a specified velocity ($V_0 \mathbf{\hat{k}}$), subjected to an external point force ($\mathbf{F}$) and a point torque ($\mathbf{T}$). (b) Schematic of the $Fe_3O_4$ nanoparticles encapsulated within the PDMS matrix \citep{peng2008magnetically}. (c) Unbounded velocity field ($\mathbf{v}_{un}$) undergoing subsequent reflections from wall (w) and sphere (s).}
\label{fig:1}
\end{figure}
Therefore, we use the method of reflections in this work to analyse the wall effects on the dynamics of an elastic particle for two cases: first, when the gravity is aligned with particle velocity and second, when the gravity is in an arbitrary direction. The particle is considered as a weakly elastic sphere that translates parallel to a rigid wall. We model the particle using Navier elasticity equations and the fluid using the Stokes equations. An external point force and a point torque are applied to constrain the motion of the particle parallel to the wall as shown in figure \ref{fig:1}(a). 

The concept of the point force applied at the centre of a particle is relevant to physical systems such as smart polymeric beads, particularly magnetoresponsive polymer beads \citep{philippova2011magnetic}. These beads typically consist of a polymer matrix embedded with magnetic nano- or microparticles (MPs), where the polymer imparts structural stability and elasticity. Such systems are used in biomedical applications—for instance, in the in vitro magnetic separation of biological cells. Non-magnetic particles can be bound to magnetically responsive beads for manipulation via magnetic fields \citep{safarik1995application,safarikova2001application,philippova2011magnetic}. The magnetic particles may be distributed either uniformly or non-uniformly within the polymer matrix \citep{kim2005magnetomicelles,peng2008magnetically,philippova2011magnetic}. The schematic of encapsulated MPs inside the polymer bead is shown in figure \ref{fig:1}(b). The average number of MPs per polymer bead can be tuned by adjusting the concentrations of the polymer and MPs \citep{kim2005magnetomicelles}.
The response of each embedded magnetic particle inside a bead to the external magnetic field can be interpreted as a localised force. Although the magnetic field is distributed over the finite region occupied by the magnetic particle, its small size relative to the bead allows this force to be effectively approximated as a point force. Such a bead can also experience localised torque when exposed to a rotating magnetic field \citep{moerland2019rotating}. In our analysis, we considered the simplest model for the bead in which a point force and a point torque are positioned at the centre of an elastic sphere. 

The problem setup in the present analysis corresponds to a resistance problem, in which the forces and the torques required to achieve a specified motion are determined. Murata solved the mobility problem for a weakly elastic sphere sedimenting under gravity in an unbounded domain \citep{murata1980deformation}. In a mobility problem, the motion of the particle is determined for a given set of forces and torques \citep{kim2013microhydrodynamics}. For a rigid particle, the results from the mobility problem can be used to derive those of the resistance problem. However, such a relationship need not hold for a deformable particle. For a deformable particle, its shape depends on the force distribution acting on it, which can differ between the mobility and resistance problems. Murata's results are for a specific force distribution that arises from the balance of gravity and hydrodynamic forces. However, the force distribution in our case is different, and hence,  we cannot use his results for our analysis in an unbounded domain. Therefore, we first analyse the motion of the particle in an unbounded domain as a resistance problem, and subsequently account for the wall effects using the method of reflections. We use a theoretical approach similar to the one used in Murata's work, employing domain perturbation and series solutions for the governing equations of both the particle and fluid. The deformed shape we obtain for the first case (gravity aligned with particle velocity) differs from that obtained in Murata's work, due to the difference in the force distribution in both problems. Our analysis of the unbounded problem in the second case (gravity is in an arbitrary direction) reveals that hydrodynamic forces acting along the translational direction are modified by body forces acting perpendicular to the motion. The wall effects on particle dynamics are analysed using the method of reflections. In the method, the disturbance velocity field generated by the particle’s motion in an unbounded domain is iteratively reflected between the wall and particle surfaces to satisfy boundary conditions on both. Reflections are carried out to obtain results up to $\textit{O}(1/H^3)$ and $\textit{O}(1/H^2)$ for the first and the second cases, respectively, assuming $\alpha$ is of the order $1/H$. 
The importance of this assumption that simplifies the analysis is discussed in section \ref{subsec:grav_parallel}. Hydrodynamic forces due to wall effects generate a lift force away from the wall. We find that the point force balancing the lift is non-zero in the first case and zero in the second. Further, we find that the point torque is zero in the first case but non-zero in the second.

The subsequent sections of this manuscript are organised as follows: Section \ref{sec:mathematical formulation} presents the mathematical formulation, the governing equations, and the boundary conditions. The results are presented for the first and second case in section \ref{subsec:grav_parallel} and \ref{subsec:grav_arb}, respectively. The results for a neutrally buoyant particle and for zero gravity are presented in section \ref{subsec:grav_parallel}. We summarise our investigation in section \ref{sec:summary}.

\section{Mathematical formulation}\label{sec:mathematical formulation}
We consider a weakly elastic spherical particle of undeformed radius ($R_0$) moving in a quiescent Newtonian fluid. The particle is translating parallel to a rigid wall at a velocity ($V_0 \mathbf{\hat{k}}$) as shown in figure \ref{fig:1}(a). The sphere centre is at a distance $h$ from the wall. An external point force ($\mathbf{F}$) and a point torque ($\mathbf{T}$) are applied at the centre to ensure that its motion is parallel to the wall. The unit vector normal to the wall ($\hat{\mathbf{n}}_w$) is aligned with the x-axis. We assume that the inertial effects in the flow are negligible compared to the viscous effects, i.e. the Reynolds number, defined as $\Rey =\rho_fV_0 R_0/ \mu \ll 1$. Here, $\rho_f$ and $\mu$ are the fluid density and viscosity, respectively.  We assume that the fluid is incompressible and is governed by the Stokes equations given by
\begin{align}
 - \mathbf{\nabla} p+\mu \nabla^2  \mathbf{v} +\rho_f \mathbf{g} ={0},  \label{eq:stokes} \\
   \mathbf{\nabla}\cdot \mathbf{v}={0}. \label{eq:continuity}
 \end{align}
In the above equations, $\mathbf{v}$ and $p$ are the velocity and pressure fields in the fluid, respectively, and $\mathbf{g}$ is the acceleration due to gravity. The particle is modelled using the Navier elasticity equations given by 
\begin{align}
  \mathbf[\lambda+G]  \mathbf{\nabla}\left( \mathbf{\nabla} \cdot  \mathbf{u}\right)+G  \mathbf{\nabla}^2  \mathbf{u} +\rho_p \mathbf{g} +\mathbf{F}\delta(\mathbf{x})+\frac{1}{2}\left[\mathbf{\nabla}\delta(\mathbf{x})\times \mathbf{T} \right] ={0}.\label{eq:elasticity}
\end{align}
Above, $\mathbf{u}$ is the displacement field in the sphere, $\lambda$ and $G$ are the Lam\'e's constants, and $\rho_p$ is the particle density. The fluid stress ($\boldsymbol{\sigma}$) is given by
\begin{align}
 \boldsymbol{\sigma}=-p \mathbf{I}+\mu \left(\nabla \mathbf{v}+[\nabla \mathbf{v}]^T\right) \label{eq:fluid-stress constitutive},
\end{align}
and the solid stress ($\boldsymbol{\tau}$) is given by
\begin{align}
   \boldsymbol{\tau}=\lambda \left(\nabla\cdot\mathbf{u}\right)\mathbf{I}+G\left(\nabla \mathbf{u}+[\nabla \mathbf{u}]^T\right).\label{eq:solid-stress constitutive}
\end{align}
The variables in the problem are non-dimensionalised using the scalings given below
\begin{align}
    \mathbf{v^*}=\frac{\mathbf{v}}{V_0};
    p^*=\frac{p R_0}{\mu V_0}; \boldsymbol{\sigma}^*=\frac{\boldsymbol{\sigma} R_0}{\mu V_0};
    \mathbf{u^*}=\frac{\mathbf{u}}{R_{0}};
    \boldsymbol{\tau}^*=\frac{\boldsymbol{\tau}}{G};H^*=\frac{h}{R_0}. \label{eq:nondimensional variables}
\end{align}
The deformability can be quantified as a measure of elastic strain when subjected to viscous stress, defined as $\alpha=\mu V_0/(G R_0)$ \citep{finney2024impact}. The parameter $\alpha$ can also be interpreted as an elastic capillary number, representing the ratio of viscous to elastic stress \citep{villone2019dynamics}, or as the ratio of viscous to elastic forces \citep{murata1980deformation,murata1981deformation,nasouri2017elastic}. The ratio of the bulk modulus ($\lambda$) to the shear modulus ($G$) of the particle is denoted by $\gamma$; $\gamma= \lambda/G$. Non-dimensional variables are indicated using the asterisk symbol in (\ref{eq:nondimensional variables}), which is omitted hereafter for brevity.

To analyse the dynamics of the particle, we express the solutions of (\ref{eq:stokes}) and (\ref{eq:elasticity}) in a spherical coordinate system ($r$,$\theta$,$\phi$), whose origin coincides with the centre of the undeformed particle. The solutions are expressed in terms of $\xi$, $\theta$, and $\phi$, where $\xi$ is the non-dimensional radial coordinate defined by $\xi=r/R_0$, $\theta$ and $\phi$ are the polar and azimuthal angles, respectively. Since the wall-particle gap width is constant, the particle deforms and reaches a steady shape whose surface radius is defined by
\begin{align}
  \xi=1+f(\theta,\phi). \label{eq:surface}
\end{align}
Here, $f$ quantifies the surface deformation and is a function of the polar and azimuthal angles. Note that in an unbounded domain, if gravity is aligned with the particle velocity, $f$ is axisymmetric about the translation axis \citep{murata1980deformation}. The function $f$ is related to the displacement field in the particle. 
The equation that relates the displacement of a material point on the deformed surface to its location on the undeformed surface is obtained in terms of $f$ as 
\begin{align}
    1=\left[1+f-u_r\vert_{\xi=1+f(\theta,\phi)}\right]^2+\left[u_\theta\vert_{\xi=1+f(\theta,\phi)}\right]^2+\left[u_\phi\vert_{\xi=1+f(\theta,\phi)}\right]^2 .\label{eq:deform_disp_relation}
\end{align}
The equation (\ref{eq:deform_disp_relation}) states that, for any point on the deformed surface, the magnitude of the difference between its current position and the displacement it experienced is equal to one. Since the particle is weakly deformable ($\alpha\ll 1$), its shape deviates slightly from the spherical shape. Noting that $\alpha$ is small, we have expanded all the variables in the problem as series in $\alpha$  and used the domain perturbation method \citep{rangasbook,garyleal} to transfer the boundary conditions from the deformed to the undeformed surface of the particle. This generates the modified boundary conditions at the undeformed surface. The series solutions to the governing equations in (\ref{eq:stokes})-(\ref{eq:elasticity})  are used to find the fields that satisfy the modified boundary conditions. To include the effect of the wall, we have used the method of reflections \citep{kim2013microhydrodynamics} and the fields at each order in $\alpha$ are expanded as series in $1/H$, the inverse of the non-dimensional wall-particle gap width. The external point force and the point torque are calculated based on the derived velocity and displacement fields at each order.  We present the series solutions to the Stokes equations and the Navier elasticity equations in the above-mentioned spherical coordinate system in sections \ref{subsec:series solution to Stokes eq} and \ref{subsec:series solution to elasticity eq}.  We explain the domain perturbation method for the case of the particle translating in an unbounded domain in section \ref{subsec:domain perturbation method}. The case of the particle translating parallel to a wall is discussed in section \ref{subsec:methodofreflections}. We summarise the steps to estimate the point force and the point torque in section \ref{subsec:solution procedure}. The results are presented in section \ref{sec:results}. 
\subsection {Series solution to Stokes equations}\label{subsec:series solution to Stokes eq}
The solutions to (\ref{eq:stokes}) and (\ref{eq:continuity}) are written in a general form below \citep{kim2013microhydrodynamics}. The radial ($v_r$), tangential ($v_\theta$) and azimuthal ($v_\phi$) velocity components are given by
 \begin{align}
v_r=\sum_{n=1}^{\infty} \sum_{m=0}^{n}& \left[\left (\frac{\xi^{-n}(n+1)}{2(2n-1)}P_n^m[A_{mn} \cos m\phi + \tilde{A}_{mn} \sin m\phi]\right)- \right.\nonumber\\& \left.
\vphantom{\frac{(n+1)}{2(2n-1)}}\left(\xi^{-n-2}(n+1)P_n^m[B_{mn} \cos m\phi +\tilde{B}_{mn} \sin m\phi]\right)\right] ,\label{eq:vr series}
\end{align}
\begin{align}
v_{\theta}= \sum_{n=1}^{\infty} \sum_{m=0}^{n}&\left[\left(\frac{\xi^{-n}(n-2)}{2n(2n-1)(2n+1) \sin \theta}\left[(n+1)(n+m)P_{n-1}^m-n(n-m+1)P_{n+1}^m\right]\right)\right.\nonumber\\&\left.\left[A_{mn} \cos m\phi + \tilde{A}_{mn} \sin m\phi\right]-\right.\nonumber\\&\left (\frac{\xi^{-n-2}}{(2n+1) \sin \theta} \left[(n+1)(n+m)P_{n-1}^m-n(n-m+1)P_{n+1}^m\right]\right)\nonumber\\&\left.[B_{mn} \cos m\phi + \tilde{B}_{mn} \sin m\phi]+\right.\nonumber\\&\left.
\left (\frac{\xi^{-n-1} m}{\sin \theta}P_n^m[-V_{mn} \sin m\phi +\tilde{V}_{mn} \cos m\phi]\right)\right],\label{eq:vtheta series}
\end{align}
\begin{align}
v_{\phi}= \sum_{n=1}^{\infty} \sum_{m=0}^{n}&\left[\left(-\frac{\xi^{-n}(n-2)m}{2n(2n-1)\sin \theta} P_n^m [-A_{mn} \sin m\phi + \tilde{A}_{mn} \cos m\phi]\right)+\right.\nonumber\\&\left.
\left (\frac{\xi^{-n-2} m}{\sin \theta} P_n^m [-B_{mn} \sin m\phi +\tilde{B}_{mn} \cos m\phi]\right)+\right.\nonumber\\&\left.
\left (\frac{\xi^{-n-1}}{(2n+1)\sin \theta}\left[(n+1)(n+m)P_{n-1}^m -n(n-m+1)P_{n+1}^m\right]\right)\right.\nonumber\\&\left.(V_{mn} \cos m\phi +\tilde{V}_{mn} \sin m\phi)\right],\label{eq:vphi series}
\end{align}
respectively, and the pressure is given by
\begin{align}
p&= \xi\left[K_{fz} \cos \theta + K_{fx} \cos \phi \sin \theta + K_{fy} \sin \theta \sin \phi\right] \nonumber \\& +\sum_{n=1}^{\infty} \sum_{m=0}^{n}\left[\xi^{-n-1}P_n^m(A_{mn} \cos m\phi + \tilde{A}_{mn} \sin m\phi)\right] .\label{eq:pressure series}
\end{align}
Here, $K_{fx}=\rho_f {R_0^2}g_x/(\mu V_0)$, $K_{fy}=\rho_f {R_0^2}g_y/(\mu V_0)$, $K_{fz}=\rho_f {R_0^2}g_z/(\mu V_0)$, and $\mathbf{g}=g_x\mathbf{\hat{i}}+g_y\mathbf{\hat{j}}+g_z\mathbf{\hat{k}}$ in the Cartesian cordinate system. Above, $P^m_n$ denotes $P^m_n(\cos \theta)$, the associated Legendre polynomial of order $m$ and degree $n$. The constants, $A_{mn}$, $\tilde{A}_{mn}$, $B_{mn}$, $\tilde{B}_{mn}$, $V_{mn}$, and $\tilde{V}_{mn}$ are determined from the no-slip boundary conditions at the surface of the particle.
\subsection {Series solution to Navier elasticity equations}\label{subsec:series solution to elasticity eq}
The solution to (\ref{eq:elasticity}) is expressed in a general form  below \citep{kushch2013micromechanics}. The radial ($u_r$), tangential ($u_{\theta}$) and azimuthal ($u_\phi$) displacement  components are given by
\begin{align}
u_r=&\frac{1}{4 \pi \xi}\left[F_{z} \cos \theta + F_{x} \cos \phi \sin \theta + F_{y} \sin \theta \sin \phi\right]-\nonumber \\&\frac{2\gamma+7}{30(\gamma+2)}\xi^2\left[K_{pz} \cos \theta + K_{px} \cos \phi \sin \theta + K_{py} \sin \theta \sin \phi\right]\,+\nonumber \\&\sum_{n=0}^{\infty} \sum_{m=0}^{n}\left[\xi^{n+1}P_n^m(B_{mn}^1 \cos m\phi +B_{mn}^0 \sin m\phi)\right] +\sum_{n=1}^{\infty} \sum_{m=0}^{n}\left[\xi^{n-1}P_n^m(C_{mn}^1 \cos m\phi +\right.\nonumber\\& \left.C_{mn}^0 \sin m\phi)\right],
\end{align}
\begin{align}
u_{\theta}=&\frac{(3 + \gamma)}{8 \pi (2 + \gamma) \xi}\left( -F_{z} \sin \theta + \cos \theta \left[F_{x} \cos \phi + F_{y} \sin \phi \right]\right)+\frac{1}{8 \pi \xi^2}\left(T_{y} \cos \phi - T_{x} \sin \phi \right)+\nonumber \\& \frac{4\gamma+9}{30(\gamma+2)}\xi^2\left[ K_{pz} \sin \theta - \cos \theta \left(K_{px} \cos \phi + K_{py} \sin \phi \right)\right]+\nonumber \\&\frac{1}{\sin \theta}\left[\sum_{n=1}^{\infty} \sum_{m=0}^{n}\left(\frac{\xi^n m(2n+1)}{n (n+1)}P_n^m [-A_{mn}^1 \sin m\phi +A_{mn}^0 \cos m\phi]\right)\right]\nonumber +\\&
\frac{1}{\sin \theta}\left[\sum_{n=1}^{\infty} \sum_{m=0}^{n}\left( \frac{\xi^{n+1} [\gamma(n+3)+(n+5)]}{(2n+1)(n+1)[\gamma\, n +(n-2)]}\left[n(n-m+1) P_{n+1}^m-(n+1)(n+m)P_{n-1}^m\right]\right)\right.\nonumber \\& \left.\vphantom{\sum_{m=0}^{n}\left(\xi^{n+1} \frac{n}{(2n+1)}\lambda_{Gf}\left[\frac{n-m+1}{n+1}P_{n+1}^m-\frac{n+m}{n}P_{n-1}^m\right]\right)}(B_{mn}^1 \cos m\phi +B_{mn}^0 \sin m\phi)\right]\nonumber +\\
&\frac{1}{\sin \theta}\left[\sum_{n=1}^{\infty} \sum_{m=0}^{n}\left( \frac{\xi^{n-1}}{(2n+1)n}\left[n(n-m+1) P_{n+1}^m-(n+1)(n+m)P_{n-1}^m\right]\right)\right.\nonumber\\& \left.\vphantom{\sum_{m=0}^{n}\left(\xi^{n+1} \frac{n}{(2n+1)}\left[\frac{n-m+1}{n+1}P_{n+1}^m-\frac{n+m}{n}P_{n-1}^m\right]\right)}(C_{mn}^1 \cos m\phi +C_{mn}^0 \sin m\phi)\right], \text{\, and}
\end{align}
\begin{align}
u_{\phi}= &\frac{(3 + \gamma)}{8 \pi (2 + \gamma) \xi} \left(F_{y} \cos \phi - F_{x} \sin \phi \right)-\frac{4\gamma+9}{30(\gamma+2)}\xi^2\left[ K_{py} \cos\phi - K_{px}\sin\phi\right]+\nonumber\\&\frac{1}{8 \pi \xi^2} \left(T_{z} \sin \theta - \cos \theta \left[T_{x} \cos \phi + T_{y} \sin \phi \right]\right)+\nonumber\\&\frac{1}{\sin \theta}\left[\sum_{n=0}^{\infty} \sum_{m=0}^{n}\left(\frac{-\xi^n}{(n+1)} \left[(n-m+1) P_{n+1}^m(A_{mn}^1 \cos m\phi +A_{mn}^0 \sin m\phi)\right]\right) + \right.\nonumber\\&\left.\sum_{n=1}^{\infty} \sum_{m=0}^{n}\left(\frac{\xi^n}{n}\left[(n+m)P_{n-1}^m(A_{mn}^1 \cos m\phi +A_{mn}^0 \sin m\phi) \right]\right)\vphantom{\sum_{m=0}^{n}\xi^n \left(-\frac{(n-m+1)}{n+1}P_{n+1}^m + \frac{n+m}{n}P_{n-1}^m\right )}\right]\nonumber + \\
&\frac{1}{\sin \theta}\left[\sum_{n=0}^{\infty} \sum_{m=0}^{n}\left( \frac{\xi^{n+1} m [\gamma(n+3)+(n+5)]}{(n+1)[\gamma \, n +(n-2)]} P_{n}^m [-B_{mn}^1 \sin m\phi +B_{mn}^0 \cos m\phi]\right)\right]+\nonumber \\
&\frac{1}{\sin \theta}\left[\sum_{n=1}^{\infty} \sum_{m=0}^{n} \left (\frac{\xi^{n-1} m}{n} P_{n}^m [-C_{mn}^1 \sin m\phi +C_{mn}^0 \cos m\phi]\right )\right]\nonumber,\label{eq:uphi displacement series}\\
\end{align}
respectively. Here, $K_{px}=\rho_p g_x R_0/G$, $K_{py}=\rho_p g_y R_0/G$, and $K_{pz}=\rho_p g_z R_0/G$. The constants, $A^1_{mn}$, $A^0_{mn}$, $B^1_{mn}$, $B^0_{mn}$, $C^1_{mn}$, and $C^0_{mn}$ are determined from the stress continuity at the particle surface. The displacement field for the point force and the point torque decays as $1/\xi$ and $1/\xi^2$, respectively as discussed in Appendix \ref{app:displacement field for point force and point torque}. The Cartesian components of the force and torque are denoted by $F_{i}$ and $T_{i}$ ($i=x,y,z$), respectively. We have done the analysis in a particle-fixed coordinate system, and the point torque arrests the rotation of the particle. The constants that correspond to translation ($C^1_{01}$, $C^1_{11}$, and $C^0_{11}$) and rotation ($A^1_{01}$, $A^1_{11}$, and $A^0_{11}$) in the above series are, therefore, set to zero.

\subsection{Domain pertubation method }\label{subsec:domain perturbation method}
In this section, we describe the domain perturbation method that is used to transfer the boundary conditions from the deformed surface of the particle to the undeformed surface as it translates in an unbounded domain. The particle deforms as it translates in the quiescent fluid. Since the particle is weakly deformable, its shape deviates slightly from the spherical shape. The deformation $f$ given in (\ref{eq:surface}) has to be determined, as a part of the solution. The velocity boundary condition (no-slip) and the stress continuity on the deformed surface are given by
\begin{equation}
  \mathbf{v}=0
  \label{eq:velocity bc},
\end{equation}
and
\begin{equation}
  \boldsymbol{\tau}.\mathbf{\hat{n}}=\alpha \, \boldsymbol{\sigma}.\mathbf{\hat{n}}, \label{eq:stress bc}
\end{equation}
respectively. Here, $\mathbf{\hat{n}}$ is the outward unit normal from the deformed surface defined by $\mathbf{\hat{n}}= \frac{\nabla S}{|\nabla S|}$, where $S= \xi-1-f(\theta,\phi)$. 
The stress boundary condition in terms of different stress components is given by
\begin{align}
  \tau_{rr}-\frac{\tau_{r\theta}}{\xi}\frac{\partial{f}}{\partial{\theta}}-\frac{\tau_{r\phi}}{\xi \sin\theta}\frac{\partial {f}}{\partial{\phi}}=\alpha\left[\sigma_{rr}-\frac{\sigma_{r\theta}}{\xi}\frac{\partial{f}}{\partial{\theta}}-\frac{\sigma_{r\phi}}{\xi \sin\theta}\frac{\partial {f}}{\partial{\phi}}\right],\label{eq:rr stress bc}
\end{align}
\begin{align}
  \tau_{r\theta}-\frac{\tau_{\theta\theta}}{\xi}\frac{\partial{f}}{\partial{\theta}}-\frac{\tau_{\theta\phi}}{\xi \sin\theta}\frac{\partial {f}}{\partial{\phi}}=\alpha\left[\sigma_{r\theta}-\frac{\sigma_{\theta\theta}}{\xi}\frac{\partial{f}}{\partial{\theta}}-\frac{\sigma_{\theta \phi}}{\xi \sin\theta}\frac{\partial {f}}{\partial{\phi}}\right],\label{eq:rtheta stress bc}
\end{align}
and
\begin{align}
  \tau_{r\phi}-\frac{\tau_{\theta\phi}}{\xi}\frac{\partial{f}}{\partial{\theta}}-\frac{\tau_{\phi\phi}}{\xi \sin\theta}\frac{\partial {f}}{\partial{\phi}}=\alpha\left[\sigma_{r\phi}-\frac{\sigma_{\theta\phi}}{\xi}\frac{\partial{f}}{\partial{\theta}}-\frac{\sigma_{\phi \phi}}{\xi \sin\theta}\frac{\partial {f}}{\partial{\phi}}\right].\label{eq:rphi stress bc}
\end{align}
Here, the $\tau_{ij}$ and $\sigma_{ij}$ with $i,j=r,\theta,\phi$ are the components of the solid and fluid stress tensors, respectively. The boundary conditions are shifted onto the undeformed surface using the domain perturbation method \citep{garyleal}. In this method, we first performed the Taylor series expansion of the boundary conditions (\ref{eq:velocity bc})-(\ref{eq:rphi stress bc}) about $\xi=1$. The regular asymptotic expansion of the variables in the problem ($\mathbf{v}$, $ \boldsymbol{\sigma}$, $p$, $\mathbf{u}$, $f$, $ \boldsymbol{\tau}$, $\rho_p$, $\mathbf{F}$, $\mathbf{T}$) are then substituted into these expanded boundary conditions. The regular expansions of the variables as series in $\alpha$ are given by
\begin{equation}
 \mathbf{v} = \mathbf{v}^{(0)}+\alpha\mathbf{v}^{(1)}+\alpha^2\mathbf{v}^{(2)}+...\, , \label{eq:vel_exp_in_alpha}
\end{equation}
\begin{equation}
 \boldsymbol{\sigma} =\boldsymbol{\sigma}^{(0)}+\alpha\boldsymbol{\sigma}^{(1)}+\alpha^2\boldsymbol{\sigma}^{(2)}+...\, ,\label{eq:fluidstress_exp_in_alpha}
\end{equation}
\begin{equation}
 p =p^{(0)}+\alpha p^{(1)}+\alpha^2 p^{(2)}+...\, ,
\end{equation}
\begin{equation}
 \mathbf{u} =\alpha\mathbf{u}^{(1)}+\alpha^2\mathbf{u}^{(2)}+...\, ,\label{eq:displacement_exp_in_alpha}
\end{equation}
\begin{equation}
f =\alpha f^{(1)}+\alpha^2 f^{(2)}+...\, ,\label{eq:deform_exp_in_alpha}
\end{equation}
\begin{equation}
 \boldsymbol{\tau} =\alpha\boldsymbol{\tau}^{(1)}+\alpha^2\boldsymbol{\tau}^{(2)}+...\, ,\label{eq:solidstress_exp_in_alpha}
\end{equation}
\begin{equation}
\rho_p =\rho_p^{(0)}+\alpha \rho_p^{(1)}+\alpha^2 \rho_p^{(2)}+...\, ,\label{eq:density_exp_in_alpha}
\end{equation}
\begin{equation}
 \mathbf{F} =\mathbf{F}^{(0)}+\alpha\mathbf{F}^{(1)}+\alpha^2\mathbf{F}^{(2)}+...\,,
\end{equation}
and
\begin{equation}
 \mathbf{T} =\mathbf{T}^{(0)}+\alpha\mathbf{T}^{(1)}+\alpha^2\mathbf{T}^{(2)}+...\, .\label{eq:torque_exp_in_alpha}
\end{equation}  
Note that the leading-order term in the series for the displacement field (\ref{eq:displacement_exp_in_alpha}), the deformation (\ref{eq:deform_exp_in_alpha}), and the solid stress (\ref{eq:solidstress_exp_in_alpha}) arising from the deformation are \textit{O}($\alpha$). 
\subsubsection{Modified velocity boundary condition on the undeformed surface ($\xi=1$)}\label{subsubsec:mod vel bc}
We have performed Taylor expansion of (\ref{eq:velocity bc}) about the undeformed surface ($\xi=1$) and substituted the velocity and deformation expansions from (\ref{eq:vel_exp_in_alpha}) and (\ref{eq:deform_exp_in_alpha}), respectively, to obtain the modified velocity boundary conditions at various orders in $\alpha$. At leading-order, the boundary conditions are given by\\
\begin{align}
  \mathbf{v}^{(0)}&=0 \,\,  \mbox{as $\xi$ =1}, \nonumber  \\
  &\rightarrow -1  \mathbf{\hat{k}} \,\, \mbox{as $\xi \rightarrow \infty$}.  
  \label{eq:vel_v0_bc}
\end{align}
At \textit{O}($\alpha$), it is \\
\begin{equation} \mathbf{v}^{(1)}+f^{(1)}\frac{\partial{\mathbf{v}^{(0)}}}{\partial{\xi}}=0,   \label{eq:vel_v1_bc}
\end{equation}
and at \textit{O}($\alpha^2$), it is \\
\begin{equation} \mathbf{v}^{(2)}+f^{(1)}\frac{\partial{\mathbf{v}^{(1)}}}{\partial{\xi}}+f^{(2)}\frac{\partial{\mathbf{v}^{(0)}}}{\partial{\xi}}+\frac{\left[f^{(1)}\right]^2}{2}\frac{\partial^2{\mathbf{v}^{(0)}}}{\partial{\xi}^2}=0 .\label{eq:vel_v2_bc}
\end{equation}
Further, the velocity fields $\mathbf{v}^{(1)}$ and $\mathbf{v}^{(2)}$  decays as $\xi \rightarrow \infty$. Note that the boundary conditions in  (\ref{eq:vel_v1_bc}) and (\ref{eq:vel_v2_bc}) are nonlinear. Although the governing equations are linear, the nonlinearity arises from applying the boundary conditions at the deformed surface whose shape is a part of the solution.
\subsubsection{Modified stress boundary conditions on the undeformed surface ($\xi=1$)}\label{subsubsec:mod stress bc}

We have performed Taylor expansion of (\ref{eq:stress bc}) about the undeformed surface ($\xi=1$) and substituted the fluid stress, deformation, and solid stress expansions from (\ref{eq:fluidstress_exp_in_alpha}), (\ref{eq:deform_exp_in_alpha}), and (\ref{eq:solidstress_exp_in_alpha}), respectively, to obtain the modified stress boundary conditions at various orders in $\alpha$.\\ 
At \textit{O}($\alpha$), the stress boundary conditions are
\begin{equation}
  \tau^{(1)}_{rr} = \sigma^{(0)}_{rr}, \label{eq:Oalpha_stressbc_r}
\end{equation}
\begin{equation}
  \tau^{(1)}_{r\theta} = \sigma^{(0)}_{r\theta} ,\label{eq:Oalpha_stressbc_theta}
\end{equation}
\begin{equation}
  \tau^{(1)}_{r\phi} = \sigma^{(0)}_{r\phi} .\label{eq:Oalpha_stressbc_phi}
\end{equation}
At \textit{O}($\alpha^2$), they are
\begin{equation}
    \tau_{rr}^{(2)}=\sigma_{rr}^{(1)}+f^{(1)} \frac{\partial}{\partial{\xi}}\left[\sigma_{rr}^{(0)}-\tau_{rr}^{(1)}\right],\label{eq:Oalpha^2_stressbc_r}
\end{equation}
\begin{equation}
    \tau_{r\theta}^{(2)}=\sigma_{r\theta}^{(1)}-f^{(1)} \frac{\partial}{\partial{\xi}}\left[\tau_{r\theta}^{(1)}-\sigma_{r\theta}^{(0)}\right]+\frac{\partial{f^{(1)}}}{\partial{\theta}}\left[\tau_{\theta \theta}^{(1)}-\sigma_{\theta \theta}^{(0)}\right] + \frac{1}{\sin\theta}\frac{\partial{f^{(1)}}}{\partial{\phi}}\left[\tau_{\theta \phi}^{(1)}-\sigma_{\theta \phi}^{(0)}\right],\label{eq:Oalpha^2_stressbc_theta}
\end{equation}
\begin{equation}
    \tau_{r\phi}^{(2)}=\sigma_{r\phi}^{(1)}-f^{(1)} \frac{\partial}{\partial{\xi}}\left[\tau_{r\phi}^{(1)}-\sigma_{r\phi}^{(0)}\right]+\frac{\partial{f^{(1)}}}{\partial{\theta}}\left[\tau_{\theta \phi}^{(1)}-\sigma_{\theta \phi}^{(0)}\right] + \frac{1}{\sin\theta}\frac{\partial{f^{(1)}}}{\partial{\phi}}\left[\tau_{\phi \phi}^{(1)}-\sigma_{\phi \phi}^{(0)}\right],\label{eq:Oalpha^2_stressbc_phi}
\end{equation}
and at \textit{O}($\alpha^3$), they are
\begin{align}
  \tau_{rr}^{(3)}=&\sigma_{rr}^{(2)}-f^{(1)} \frac{\partial}{\partial{\xi}}\left[\tau_{rr}^{(2)}-\sigma_{rr}^{(1)}\right]-f^{(2)} \frac{\partial}{\partial{\xi}}\left[\tau_{rr}^{(1)}-\sigma_{rr}^{(0)}\right]-\frac{\left[f^{(1)}\right]^2}{2}\frac{\partial^2}{\partial{\xi}^2}\left[\tau_{rr}^{(1)}-\sigma_{rr}^{(0)}\right]+\nonumber\\&  \frac{\partial f^{(2)}}{\partial{\theta}}\left[\tau_{r\theta}^{(1)}-\sigma_{r\theta}^{(0)}\right]+\frac{\partial f^{(1)}}{\partial{\theta}}\left[\tau_{r\theta}^{(2)}-\sigma_{r\theta}^{(1)}\right]+ f^{(1)}\frac{\partial f^{(1)}}{\partial{\theta}}\frac{\partial}{\partial{\xi}}\left[\tau_{r\theta}^{(1)}-\sigma_{r\theta}^{(0)}\right]+\nonumber\\&\frac{1}{\sin\theta}\left(\frac{\partial f^{(2)}}{\partial{\phi}}\left[\tau_{r\phi}^{(1)}-\sigma_{r\phi}^{(0)}\right]+\frac{\partial f^{(1)}}{\partial{\phi}}\left[\tau_{r\phi}^{(2)}-\sigma_{r\phi}^{(1)}\right]+ f^{(1)}\frac{\partial f^{(1)}}{\partial{\phi}}\frac{\partial}{\partial{\xi}}\left[\tau_{r\phi}^{(1)}-\sigma_{r\phi}^{(0)}\right]\right)-\nonumber\\& f^{(1)}\frac{\partial f^{(1)}}{\partial{\theta}}\left[\tau_{r\theta}^{(1)}-\sigma_{r\theta}^{(0)}\right]-\frac{1}{\sin\theta}\left(f^{(1)}\frac{\partial f^{(1)}}{\partial{\phi}}\left[\tau_{r\phi}^{(1)}-\sigma_{r\phi}^{(0)}\right]\right),\label{eq:Oalpha^3_stressbc_r}
\end{align}
\begin{align}
  \tau_{r\theta}^{(3)}=&\sigma_{r\theta}^{(2)}-f^{(1)} \frac{\partial}{\partial{\xi}}\left[\tau_{r\theta}^{(2)}-\sigma_{r\theta}^{(1)}\right]-f^{(2)} \frac{\partial}{\partial{\xi}}\left[\tau_{r\theta}^{(1)}-\sigma_{r\theta}^{(0)}\right]-\frac{\left[f^{(1)}\right]^2}{2}\frac{\partial^2}{\partial{\xi}^2}\left[\tau_{r\theta}^{(1)}-\sigma_{r\theta}^{(0)}\right]+\nonumber\\&  \frac{\partial f^{(2)}}{\partial{\theta}}\left[\tau_{\theta\theta}^{(1)}-\sigma_{\theta\theta}^{(0)}\right]+\frac{\partial f^{(1)}}{\partial{\theta}}\left[\tau_{\theta\theta}^{(2)}-\sigma_{\theta\theta}^{(1)}\right]+ f^{(1)}\frac{\partial f^{(1)}}{\partial{\theta}}\frac{\partial}{\partial{\xi}}\left[\tau_{\theta\theta}^{(1)}-\sigma_{\theta\theta}^{(0)}\right]+\nonumber\\&\frac{1}{\sin\theta}\left(\frac{\partial f^{(2)}}{\partial{\phi}}\left[\tau_{\theta\phi}^{(1)}-\sigma_{\theta\phi}^{(0)}\right]+\frac{\partial f^{(1)}}{\partial{\phi}}\left[\tau_{\theta\phi}^{(2)}-\sigma_{\theta\phi}^{(1)}\right]+ f^{(1)}\frac{\partial f^{(1)}}{\partial{\phi}}\frac{\partial}{\partial{\xi}}\left[\tau_{\theta\phi}^{(1)}-\sigma_{\theta\phi}^{(0)}\right]\right)-\nonumber\\& f^{(1)}\frac{\partial f^{(1)}}{\partial{\theta}}\left[\tau_{\theta\theta}^{(1)}-\sigma_{\theta\theta}^{(0)}\right]-\frac{1}{\sin\theta}\left(f^{(1)}\frac{\partial f^{(1)}}{\partial{\phi}}\left[\tau_{\theta\phi}^{(1)}-\sigma_{\theta\phi}^{(0)}\right]\right),\label{eq:Oalpha^3_stressbc_theta}
\end{align}
\begin{align}
  \tau_{r\phi}^{(3)}=&\sigma_{r\phi}^{(2)}-f^{(1)} \frac{\partial}{\partial{\xi}}\left[\tau_{r\phi}^{(2)}-\sigma_{r\phi}^{(1)}\right]-f^{(2)} \frac{\partial}{\partial{\xi}}\left[\tau_{r\phi}^{(1)}-\sigma_{r\phi}^{(0)}\right]-\frac{\left[f^{(1)}\right]^2}{2}\frac{\partial^2}{\partial{\xi}^2}\left[\tau_{r\phi}^{(1)}-\sigma_{r\phi}^{(0)}\right]+\nonumber\\&  \frac{\partial f^{(2)}}{\partial{\theta}}\left[\tau_{\theta\phi}^{(1)}-\sigma_{\theta\phi}^{(0)}\right]+\frac{\partial f^{(1)}}{\partial{\theta}}\left[\tau_{\theta\phi}^{(2)}-\sigma_{\theta\phi}^{(1)}\right]+ f^{(1)}\frac{\partial f^{(1)}}{\partial{\theta}}\frac{\partial}{\partial{\xi}}\left[\tau_{\theta\phi}^{(1)}-\sigma_{\theta\phi}^{(0)}\right]+\nonumber\\&\frac{1}{\sin\theta}\left(\frac{\partial f^{(2)}}{\partial{\phi}}\left[\tau_{\phi\phi}^{(1)}-\sigma_{\phi\phi}^{(0)}\right]+\frac{\partial f^{(1)}}{\partial{\phi}}\left[\tau_{\phi\phi}^{(2)}-\sigma_{\phi\phi}^{(1)}\right]+ f^{(1)}\frac{\partial f^{(1)}}{\partial{\phi}}\frac{\partial}{\partial{\xi}}\left[\tau_{\phi\phi}^{(1)}-\sigma_{\phi\phi}^{(0)}\right]\right)-\nonumber\\& f^{(1)}\frac{\partial f^{(1)}}{\partial{\theta}}\left[\tau_{\theta\phi}^{(1)}-\sigma_{\theta\phi}^{(0)}\right]-\frac{1}{\sin\theta}\left(f^{(1)}\frac{\partial f^{(1)}}{\partial{\phi}}\left[\tau_{\phi\phi}^{(1)}-\sigma_{\phi\phi}^{(0)}\right]\right).\label{eq:Oalpha^3_stressbc_phi}
\end{align}
The deformation of the surface at different orders $f^{(1)}$ and $f^{(2)}$ appearing in the modified boundary conditions are obtained by expanding (\ref{eq:deform_disp_relation}) using Taylor series about $\xi=1$ and substituting (\ref{eq:displacement_exp_in_alpha}) and (\ref{eq:deform_exp_in_alpha}).
At \textit{O}($\alpha$), $f^{(1)}$ is given by
\begin{equation}
    f^{(1)}=\left.u_r^{(1)}\right \vert_{\xi=1}. \label{eq:f1_Expression}
\end{equation}
At \textit{O}($\alpha^2$), $f^{(2)}$ is given by
\begin{align}
    f^{(2)}=\left(u_r^{(2)} + f^{(1)} \frac{\partial{u_r^{(1)}}}{\partial{\xi}} -\frac{\left[u_\theta^{(1)}\right]^2 + \left[u_\phi^{(1)}\right]^2}{2}\right)\left. \vphantom{\frac{\left[u_\theta^{(1)}\right]^2}{2}} \right\vert_{\xi=1}. \label{eq:f2_Expression}
\end{align}
The modified boundary conditions, together with the series solutions described in the previous sections, are used to calculate all the variables defined in (\ref{eq:vel_exp_in_alpha})-(\ref{eq:torque_exp_in_alpha}) in the unbounded domain. 

\subsection{Method of reflections }\label{subsec:methodofreflections}
To capture the effect of the wall on the dynamics of the translating particle, we have employed the method of reflections \citep{kim2013microhydrodynamics}.  The unbounded velocity field ($\mathbf{v}_{un}$) doesn't satisfy the no-slip boundary condition at the wall (see figure \ref{fig:1}(c)). The velocity field is alternatively reflected from the wall ($\mathbf{v}_{w_k}$) and the sphere ($\mathbf{v}_{s_k}$), and the reflected fields are added to the unbounded velocity field, to satisfy the boundary conditions at both the surfaces to the desired accuracy. The desired accuracy can be quantified in algebraic powers of $1/H$. Therefore, the reflected velocity fields are expanded as series in $1/H$. The total velocity field at each order in $\alpha^n$ ($n=0,1,2..$) is written as
\begin{align}
 \mathbf{v}^{(n)} = \mathbf{v}^{(n)}_{un}+&\frac{{\mathbf{v}}^{(n,\scalebox{0.6}{$H$})}_{w_1}}{H}+\frac{{\mathbf{v}}^{(n,\scalebox{0.6}{$H^2$})}_{w_1}}{H^2}+...+\frac{\mathbf{v}^{(n,\scalebox{0.6}{$H$})}_{s_1}}{H}+ \frac{\mathbf{v}^{(n,\scalebox{0.6}{$H^2$})}_{w_2}}{H^2}+\frac{\mathbf{v}^{(n,\scalebox{0.6}{$H^3$})}_{w_2}}{H^3}+...+\frac{\mathbf{v}^{(n,\scalebox{0.6}{$H^2$})}_{s_2}}{H^2} +\nonumber\\&\frac{\mathbf{v}^{(n,\scalebox{0.6}{$H^3$})}_{w_3}}{H^3}+\frac{\mathbf{v}^{(n,\scalebox{0.6}{$H^4$})}_{w_3}}{H^4}+...+\frac{\mathbf{v}^{(n,\scalebox{0.6}{$H^3$})}_{s_3}}{H^3}+
 \frac{\mathbf{v}^{(n,\scalebox{0.6}{$H^4$})}_{w_4}}{H^4}+... \, .
\end{align}
Here, the subscripts $()_{un}$ and $()_{w_k}$ denote the unbounded field and the $k^{th}$ reflected field from the wall, respectively. The subscript $()_{s_k}$ indicates that the velocity field includes the effect of $k^{th}$ reflection from the sphere. The superscript $(n,H^p)$ indicates $1/H^p$ term in the series at \textit{O}($\alpha^n$). Similar to the velocity field, the fluid stress $\boldsymbol{\sigma}^{(n)}$ and pressure $p^{(n)}$ are also expanded as series in $1/H$. The displacement $\mathbf{u}^{(n)}$ at each order in $\alpha^n(n=1,2,...)$ is expanded as a series in $1/H$ as
\begin{align}
 \mathbf{u}^{(n)} = \mathbf{u}^{(n)}_{un}+\frac{\mathbf{u}^{(n,\scalebox{0.6}{$H$})}_{s_1}}{H}+\frac{\mathbf{u}^{(n,\scalebox{0.6}{$H^2$})}_{s_2}}{H^2}+\frac{\mathbf{u}^{(n,\scalebox{0.6}{$H^3$})}_{s_3}}{H^3}+...\, .
\end{align}
Similar to the displacement field, the solid stress $\boldsymbol{\tau}^{(n)}$, deformation $ f^{(n)}$, point force $\mathbf{F}^{(n)}$ and point torque $\mathbf{T}^{(n)}$ are expanded as series in $1/H$. We have derived the results till \textit{O}($1/H^3$), which needs three reflections from the wall in the first case where gravity is aligned with the particle velocity. In the second case where gravity is in an arbitrary direction, the results are derived till \textit{O}($1/H^2$), which needs two reflections from the wall.

\subsection{Solution procedure}\label{subsec:solution procedure}
Below, we describe the steps involved in calculating the point force and the point torque required to make the particle translate parallel to the wall. Steps (i)-(iv) are for an unbounded domain, and (v)-(ix) are for capturing the wall effects using the method of reflections.
\begin{enumerate}
 \item []
    \item First, we calculated the fields ($\mathbf{v}^{(0)}_{un},p^{(0)}_{un}$) surrounding the particle in an unbounded domain. At \textit{O}(1), the constants in (\ref{eq:vr series})-(\ref{eq:vphi series}) are determined by applying the velocity boundary condition in  (\ref{eq:vel_v0_bc}) and the corresponding fluid stress field  $\boldsymbol{\sigma}_{un}^{(0)}$ is calculated.
     \item []
    \item Using stress continuity in (\ref{eq:Oalpha_stressbc_r})-(\ref{eq:Oalpha_stressbc_phi}), $\boldsymbol{\tau}_{un}^{(1)}$, $\mathbf{u}_{un}^{(1)}$,  $\mathbf{F}_{un}^{(0)}$, $\mathbf{T}_{un}^{(0)}$, and $f_{un}^{(1)}$ are calculated.
    \item [] 
    \item We then used (\ref{eq:vel_v1_bc}) to determine $\mathbf{v}_{un}^{(1)}$ and in turn calculate $\boldsymbol{\sigma}_{un}^{(1)}$.
    \item[]   
    \item{ Above steps (ii and iii) are repeated at the next orders in $\alpha$ to determine the velocity and the stress fields in the fluid, the external forces and the external torques till \textit{O}($\alpha^2$), and the displacement and stress fields in the solid till \textit{O}($\alpha^3$)}.
    \item[]
    \item{} To capture the wall effects, the unbounded velocity and pressure fields till \textit{O}($\alpha^2$) are reflected from the wall, to determine the first wall reflected fields at various orders given by $\left[\frac{\mathbf{v}^{(0,\scalebox{0.6}{$H$})}_{w_1}}{H}+\alpha \frac{\mathbf{v}^{(1,\scalebox{0.6}{$H$})}_{w_1}}{H}+\alpha^2 \frac{\mathbf{v}^{(2,\scalebox{0.6}{$H$})}_{w_1}}{H}\right]$, $\left[\frac{\mathbf{v}^{(0,\scalebox{0.6}{$H^2$})}_{w_1}}{H^2}+\alpha \frac{\mathbf{v}^{(1,\scalebox{0.6}{$H^2$})}_{w_1}}{H^2}\right]$,    
 and $\frac{{\mathbf{v}}^{(0,\scalebox{0.6}{$H^3$})}_{w_1}}{H^3}$.
   
    \item[]
    \item{} The velocity fields ${\mathbf{v}}^{(n,\scalebox{0.6}{$H$})}_{w_1}$(with n=0,1,2) are reflected from the sphere to determine the first sphere reflected fields given by  $\left[\frac{\mathbf{v}^{(0,\scalebox{0.6}{$H$})}_{s_1}}{H}+\alpha \frac{\mathbf{v}^{(1,\scalebox{0.6}{$H$})}_{s_1}}{H}+\alpha^2 \frac{\mathbf{v}^{(2,\scalebox{0.6}{$H$})}_{s_1}}{H}\right]$ and the displacement fields given by $\left[\alpha \frac{\mathbf{u}^{(1,\scalebox{0.6}{$H$})}_{s_1}}{H}+\alpha^2 \frac{\mathbf{u}^{(2,\scalebox{0.6}{$H$})}_{s_1}}{H}\right]$ using the boundary conditions at the sphere surface. We followed the same steps as those described for the unbounded domain to calculate the higher-order term in $\alpha$.
    \item[]
    \item{}The velocity fields ${\mathbf{v}}^{(n,\scalebox{0.6}{$H$})}_{s_1}$ (with n=0,1) are reflected from the wall to determine the second wall reflected fields given by  $\left[\frac{\mathbf{v}^{(0,\scalebox{0.6}{$H^2$})}_{w_2}}{H^2}+\alpha \frac{\mathbf{v}^{(1,\scalebox{0.6}{$H^2$})}_{w_2}}{H^2}\right] $ and $ \frac{\mathbf{v}^{(0,\scalebox{0.6}{$H^3$})}_{w_2}}{H^3}$. All the \textit{O}($1/H^2$) terms from the first and second wall reflections are used to determine the second sphere reflected fields given by $\left[\frac{\mathbf{v}^{(0,\scalebox{0.6}{$H^2$})}_{s_2}}{H^2}+\alpha \frac{\mathbf{v}^{(1,\scalebox{0.6}{$H^2$})}_{s_2}}{H^2}\right]$ and the displacement field given by $\left[\alpha \frac{\mathbf{u}^{(1,\scalebox{0.6}{$H^2$})}_{s_2}}{H^2}\right]$ using the boundary conditions at the sphere surface.
    \item[]
    \item{} The velocity field ${\mathbf{v}}^{(0,\scalebox{0.6}{$H^2$})}_{s_2}$  is reflected from the wall to determine the third wall reflected field given by $\frac{{\mathbf{v}}^{(0,\scalebox{0.6}{$H^3$})}_{w_3}}{H^3}$. All the \textit{O}($1/H^3$) terms in the first three wall reflections are used to calculate the third sphere reflected field given by $\frac{{\mathbf{v}}^{(0,\scalebox{0.6}{$H^3$})}_{s_3}}{H^3}$.
    \item[]
    \item{}At each of the reflections from the sphere, we also determine the point force and the point torque at different orders.
\end{enumerate}
 
\section{Results}\label{sec:results}
Below, we discuss the results of our analysis. The discussion is presented for two cases: 1) gravity aligned with the particle velocity in section \ref{subsec:grav_parallel} and  2) gravity in an arbitrary direction in section \ref{subsec:grav_arb}.
\subsection{Gravity aligned with particle velocity}\label{subsec:grav_parallel}
First, we discuss the results for the case of translation of the particle along the direction of gravity ($\mathbf{g}=g_z \mathbf{\hat{k}}$) in an unbounded domain. The leading-order velocity field [$\mathbf{v}_{un}^{(0)}$] is obtained by applying  (\ref{eq:vel_v0_bc}) to determine the constants in the series solution in section \ref{subsec:series solution to Stokes eq}. The field is the same as that in a uniform flow past a rigid sphere of radius $R_0$ in an unbounded domain \citep{guazzelli2011physical}. The stress field at \textit{O}($\alpha$) in the solid is calculated using the boundary conditions (\ref{eq:Oalpha_stressbc_r})-(\ref{eq:Oalpha_stressbc_phi}), and the corresponding displacement components are obtained as
\begin{align}
    &{u^{(1)}_{un,r}}=\frac{1}{\xi}C_1 \cos\theta+\xi^2 C_2\cos\theta,\\
    &u^{(1)}_{un,\theta}=\frac{1}{\xi}\left[-\frac{3+\gamma}{4+2\gamma}C_1 \right]\sin\theta+\xi^2\left[\frac{(3 + 2\gamma) (2 K_{fz} - 9 (1 + \gamma)) - K_{fz} \gamma \tilde{\rho}}{(-1 + \gamma) (9 - 2 K_{fz} + 9 \gamma) + K_{fz} (4 + 5 \gamma) \tilde{\rho}} C_2 \right]\sin\theta.
\end{align}
Here,
\begin{align}
    &C_1=\frac{1}{6} \left( 9 - 2 K_{fz} ( -1 + \tilde{\rho} ) \right), \nonumber\\
    &C_2=-\frac{-9 + 9 \gamma^2 + K_{fz} (2 + 4 \tilde{\rho} + \gamma (-2 + 5 \tilde{\rho}))}{6 (2 + \gamma) (2 + 3 \gamma)}.\label{eq:c2 expression gravity aligned}
\end{align}
Above, $\tilde{\rho}=\rho^{(0)}_p/\rho_f$ and $K_{fz}=\rho_f {R_0^2}g_z/(\mu V_0)$, ratio of gravitational to the viscous forces. The surface deformation $f^{(1)}_{un}$ is calculated from (\ref{eq:f1_Expression}) and is obtained as
\begin{align}
f^{(1)}_{un}=C_3 \cos\theta, \label{eq:f1 unbounded gravity aligned}
\end{align}
where $C_3=C_1+C_2$. Using the boundary condition (\ref{eq:vel_v1_bc}), the velocity components and pressure in the fluid at \textit{O}($\alpha$) are obtained as
\begin{align}
&v^{(1)}_{un,r}= \left[\frac{1}{\xi^2}-\frac{1}{\xi^4}\right]\frac{3 \,C_3}{4}(3\cos^2\theta-1), \label{eq:vr1 unbounded}\\
&v^{(1)}_{un,\theta}= -\frac{1}{\xi^4}\left[\frac{3\,C_3}{2}\right]\cos\theta \sin\theta, \\
&v^{(1)}_{un,\phi}= 0, \\
&p^{(1)}_{un}=\frac{1}{\xi^3}\left[\frac{3\,C_3}{2}\right](3\cos^2\theta-1).
\end{align}
The stress field in the solid at \textit{O}($\alpha^2$) is calculated using the boundary conditions (\ref{eq:Oalpha^2_stressbc_r})-(\ref{eq:Oalpha^2_stressbc_phi}) and the corresponding displacement components are obtained as
\begin{align}
&u^{(2)}_{un,r}=\xi\left[A_1+A_2 \cos^2\theta\right]+\xi^3 A_3(1-3\cos^2\theta),  \\
&u^{(2)}_{un,\theta}= -\xi A_2\cos\theta\sin\theta+\xi^3\left[\frac{(7+5\gamma)}{\gamma}A_3\right]\sin\theta\cos\theta.
\end{align}
Here,
\begin{align}
&A_1=\left[\frac{-168 + \gamma (-48 + 3 \gamma (152 + 97 \gamma) - 2 K_{fz} (48 + \gamma (98 + 45 \gamma)) ( - 1 + \tilde{\rho}))}{4 (2 + \gamma) (2 + 3 \gamma) (14 + 19 \gamma)}\right]C_3, \nonumber\\
&A_2=\left[\frac{-336 - 3 \gamma (214 + 97 \gamma) + 2 K_{fz} (4 + 5 \gamma) (14 + 9 \gamma) (-1 + \tilde{\rho})}{4 (2 + \gamma) (14 + 19 \gamma)} \right]C_3, \nonumber\\
&A_3=\left[\frac{2 \gamma (1 + \gamma) (-9 + 2 K_{fz} (-1 + \tilde{\rho}))}{(2 + \gamma) (14 + 19 \gamma)}\right]C_3.
\end{align}
The surface deformation $f^{(2)}_{un}$ is calculated from (\ref{eq:f2_Expression}) and is obtained as
\begin{align}
f^{(2)}_{un}=  D_1+D_2 \cos^2\theta -\frac{1}{2}D_3^2 \sin^2\theta. \label{eq:f2 unbounded gravity aligned}
\end{align}
Here,
\begin{align}
   &D_1=A_1+A_3, \nonumber\\
   &D_2=A_2 - 3 A_3 - C_3 (-3 C_2 + C_3),\nonumber\\
   &D_3=\frac{(C_2 - C_3) (3 + \gamma)}{2 (2 + \gamma)} + \frac{C_2 \left( (3 + 2 \gamma) (2 K_{fz} - 9 (1 + \gamma)) - K_{fz} \gamma \tilde{\rho} \right)}{(-1 + \gamma) (9 - 2 K_{fz} + 9 \gamma) + K_{fz} (4 + 5 \gamma) \tilde{\rho}}.
\end{align}
Using (\ref{eq:vel_v2_bc}), we calculated $\mathbf{v}^{(2)}_{un}$ and then derived the stress field $\boldsymbol{\tau}^{(3)}_{un}$ using  the boundary conditions  (\ref{eq:Oalpha^3_stressbc_r})-(\ref{eq:Oalpha^3_stressbc_phi}). The expressions for components of $\mathbf{v}^{(2)}_{un}$ are provided in (\ref{eq:app v2r unbounded}) and (\ref{eq:app v2theta unbounded}) of Appendix \ref{app:gravity aligned}. The point force  till \textit{O}($\alpha^2$) in dimensional form is obtained as
\begin{align}
    \mathbf{F}_{un}&= 6\pi\mu V_0 R_{0}\left(\left[1+\frac{2 K_{fz}}{9}\left(1-\tilde{\rho} \right)\right]+\alpha^2 \left[\vphantom{\frac{2 K_{fz}}{9}}E_1+K_{fz}^3(E_2 I_1)+K_{fz}^2( E_2 E_3 I_2)+\right.\right. \nonumber\\& \left.\left.K_{fz}(E_4 E_2 I_3)\vphantom{\frac{2 K_{fz}}{9}}\right]\right)\mathbf{\hat{k}} \label{eq:unbound force gravity aligned}\,.
\end{align}
The expressions for $E_1$, $E_2$, $E_3$, $E_4$, $I_1$, $I_2$, and $I_3$ in the equation are given in (\ref{appeq:E1 E2 expression})-(\ref{appeq:I3 expression}) of Appendix \ref{app:gravity aligned}. The leading order force in the equation reduces to the sum of Stokes drag and the buoyancy force on a rigid sphere. The particle should not experience a hydrodynamic lift force or a torque due to the symmetry of the flow field about the $z$-axis in the unbounded domain. Consistently, the point force at \textit{O}($\alpha^2$) in the equation acts only along gravity ($z$-axis) and the point torque is also zero. The point force at \textit{O}($\alpha^2$) balances the hydrodynamic forces arising from the change in the particle shape and the buoyancy force. The change in the buoyancy force at \textit{O}($\alpha^2$) arises from compressibility-induced variations in the particle’s volume and density. 

\begin{figure}
  \centerline{\includegraphics[width=1\textwidth]{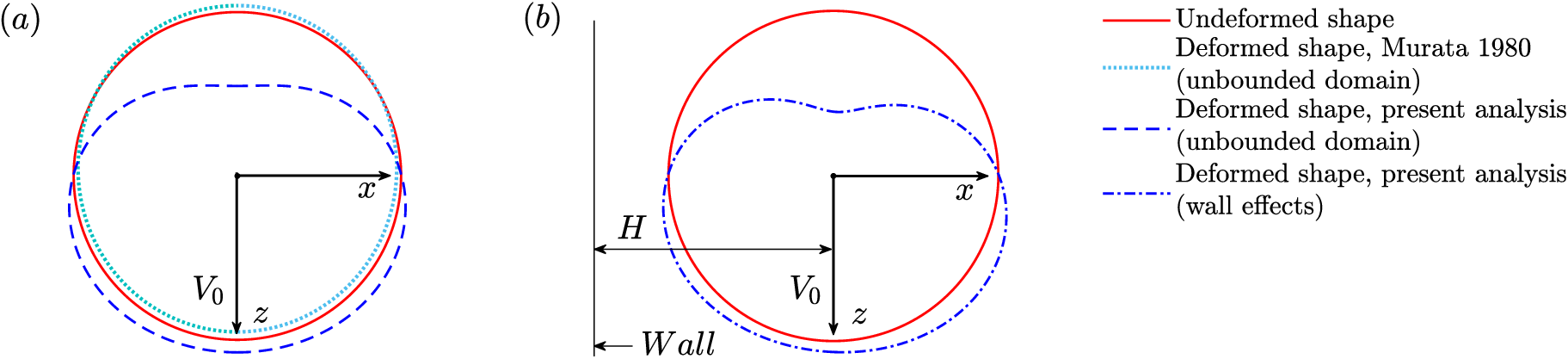}}
  \caption{ The shape of the deformed sphere on $xz$ plane at  $\alpha=0.2$, $\tilde{\rho}=1.05$, $K_{fz}=0.5$ and $\gamma=2$ in the (\textit{a}) unbounded domain and (\textit{b}) in the presence of wall at $H=2.5$.}
\label{fig:def_unbound_bound}
\end{figure}

To analyse the shape of the deformed particle, we have plotted it on $y=0$ plane as shown in figure \ref{fig:def_unbound_bound}(a). 
As is evident in the figure, the shape is symmetric about the $z$-axis. The dip at the top is due to the effect of the point force applied at the centre of the particle.  The shape we obtain differs from that obtained by Murata \cite{murata1980deformation}, also shown in the figure. Murata calculated the velocity of a weakly elastic spherical particle sedimenting due to gravity. He showed that the particle deforms to a prolate spheroid. The shape difference between ours and Murata's work arises from the varying force distributions: our problem considers equilibrium between external, gravitational, and hydrodynamic forces, whereas Murata's formulation considers only gravitational and hydrodynamic forces. Consequently, the hydrodynamic forces acting on the particle differ between the two problems. Additionally, our approach accounts for the density change due to the compressibility of the particle, an aspect not considered in Murata’s calculation for the terminal velocity, although he models the particle as a compressible one. 

\begin{figure}
  \centerline{\includegraphics[width=0.9\textwidth]{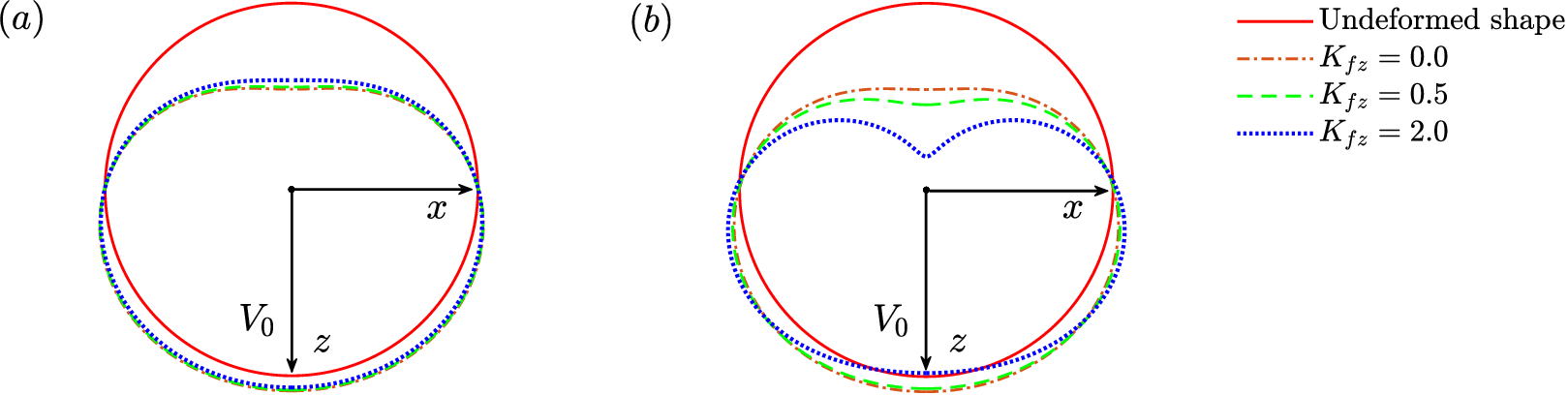}}
  \caption{Variation in the shape of the deformed sphere in the unbounded domain as a function of $K_{fz}$ at $\gamma=2$, $\alpha=0.2$ when (a) the particle is denser than fluid, $\tilde{\rho}=1.05$ and (b) when the particle is lighter than the fluid, $\tilde{\rho}=0.05$.}
\label{fig:gravity strength variatio unbounded}
\end{figure}

\begin{figure}
  \centerline{\includegraphics[width=0.85\textwidth]{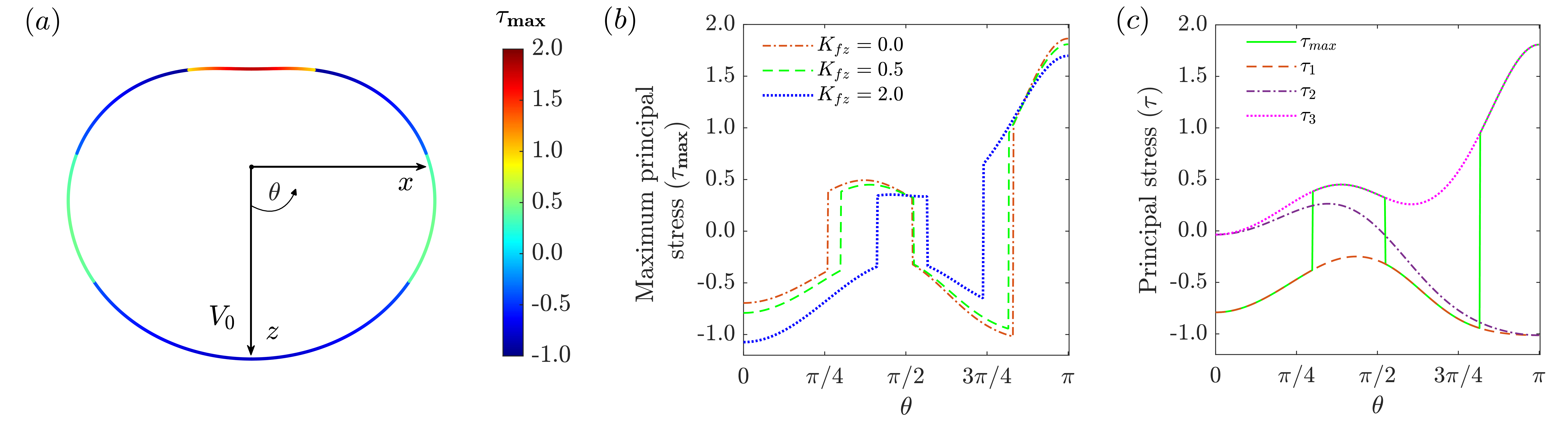}}
  \caption{Variation of the maximum principal stress (based on its absolute value) along the particle surface in the unbounded domain (a) for $K_{fz}=0.5$ and (b) for different values of $K_{fz}$. (c) Variation of the principal stresses ($\tau_1$, $\tau_2$, and $\tau_3$) and $\tau_{max}$ for $K_{fz}=0.5$, along the particle surface, plotted at $\gamma=2$, $\alpha=0.2$, and $\tilde{\rho}=1.05$ for $\theta$ varying from $0$ to $\pi$.}
\label{fig:scatterplot and gravity strength variation unbounded}
\end{figure}

The deformed shape of the particle depends on the ratio of the particle density to the ambient fluid ($\tilde{\rho}$). For a particle heavier than the surrounding fluid, as $K_{fz}$ [defined as $\rho_f {R_0^2}g_z/(\mu V_0)$ below (\ref{eq:pressure series})] increases, the external force required to balance hydrodynamic drag diminishes, resulting in a reduction in the dip in the particle's upper section as shown in figure \ref{fig:gravity strength variatio unbounded}(a). Conversely, for a particle lighter than the ambient fluid, an increase in  $K_{fz}$  necessitates a larger external force to sustain translation, which in turn enhances the dip in the particle's upper section as shown in figure \ref{fig:gravity strength variatio unbounded}(b).
Figure \ref{fig:scatterplot and gravity strength variation unbounded}(a) illustrates the variation of the maximum principal stress (based on its absolute value), denoted by $\tau_{max}$, along the surface of a heavier particle for $K_{fz}=0.5$. In figure \ref{fig:scatterplot and gravity strength variation unbounded}(b), $\tau_{max}$ is plotted against $\theta$ for different values of $K_{fz}$. The maximum principal stress is tensile in nature near $\theta=\pi$, the location of the dip in the deformed shape. As $K_{fz}$ increases, the point force diminishes, leading to a reduction in the maximum principal stress near $\theta = \pi$. This trend aligns with the less pronounced dip seen in figure \ref{fig:gravity strength variatio unbounded}(a). The observed jumps in figure \ref{fig:scatterplot and gravity strength variation unbounded}(b) occur at locations where the nature of the maximum principal stress changes, from compressive to tensile or vice-versa for different values of $K_{fz}$. To see the jumps, we have plotted the variation of principal stresses ($\tau_1$, $\tau_2$, and $\tau_3$) and $\tau_{max}$ against $\theta$, corresponding to $K_{fz} = 0.5$ along the particle surface in figure \ref{fig:scatterplot and gravity strength variation unbounded}(c), as $\theta$ varies from $0$ to $\pi$. As is evident from the figure, the principal stresses remain continuous, whereas discontinuities (jumps) appear in the plot of $\tau_{max}$ due to the use of the true value of maximum absolute principal stress. Due to the applied point force at the origin, the lower-most surface of the particle experiences compressive stress, while the upper-most surface is under tensile stress. Along the surface, the stress undergoes transition from compressive to tensile, then tensile to compressive, and finally from compressive back to tensile, as $\theta$ varies from $0$ to $\pi$.

Now we discuss the case of the particle translating next to a wall. In the presence of the wall, the calculations are carried out using the method of reflections described in section \ref{subsec:methodofreflections}. The velocity and displacement fields are calculated at orders $1/H$, $\alpha/H$, $\alpha^2/H$, $1/H^2$, $\alpha/H^2$, and $1/H^3$. The surface deformation till the order considered is obtained as
\begin{align}
f&= f_{un}+\alpha \left[ \frac{B_1 \cos\theta}{H}\left(1+\frac{9}{16 H}\right)+\frac{45 \cos\phi \sin 2\theta}{64 H^2}\right]+\alpha^2\left[\left(\frac{B_2}{H}+K_{fz}\frac{B_3}{H}\right)+\right. \nonumber\\& \left.\left(\frac{B_4}{H}+ K_{fz}\frac{B_5}{H}\right)\cos2\theta\right].\label{eq:wall induced deformation gravity aligned}
\end{align}
The expressions for $B_1$, $B_2$, $B_3$, $B_4$, and $B_5$ in the equation are given in (\ref{appeq:B1 B2 expression})-(\ref{appeq:B5 expression}) of Appendix \ref{app:gravity aligned}. The  deformation in the unbounded domain $f_{un}=\alpha f^{(1)}_{un}+\alpha^2 f^{(2)}_{un}$, where $f^{(1)}_{un}$ and $f^{(2)}_{un}$ are presented in (\ref{eq:f1 unbounded gravity aligned})  and (\ref{eq:f2 unbounded gravity aligned}), respectively. Due to the presence of the wall, the deformed shape is not symmetric about the $z$-axis and is a function of $\phi$ at \textit{O}($\alpha/H^2$). This contrasts with the unbounded case, where the deformed shape remains independent of $\phi$. We have plotted the shape of the particle in the $xz$ plane passing through the origin in figure \ref{fig:def_unbound_bound}(b) to illustrate the asymmetry. The dash-dot curve is plotted using (\ref{eq:surface}) and the shape is not symmetric about the $z$-axis.

The point force on the particle till the order considered is obtained as
\begin{align}
    \mathbf{F}&=\mathbf{F}_{un}+6\pi\mu V_0 R_{0}\left(\left[\frac{9 }{16 H} +\frac{81}{256 H^2}+\frac{217}{4096 H^3}\right] \mathbf{\hat{k}}-\frac{\alpha}{H^2} \left[\frac{9}{64}\right] \mathbf{\hat{i}}+\frac
    {\alpha^2}{H} \left[\frac{9E_1}{4}+K_{fz}^2 E_2I_4 +\right.\right. \nonumber\\& \left.\left.\vphantom{\left[\frac{9 }{16 H} +\frac{81}{256 H^2}+\frac{217}{4096 H^3}\right]} K_{fz}I_5\right]\mathbf{\hat{k}}\right) \label{eq:wall force gravity aligned}.
\end{align}
The expressions for $I_4$ and $I_5$ are given in (\ref{appeq:I4 expression}) and (\ref{appeq:I5 expression}) of Appendix \ref{app:gravity aligned}. The point force due to wall effects at the leading order balances the buoyancy force and the hydrodynamic force, derived in the literature for the case of a rigid sphere \citep{faxen1923bewegung,goldman1967slow}. It is important to note that the elastic effects lead to a hydrodynamic lift away from the wall, which is balanced by the point force at \textit{O}($\alpha/H^2$) in the equation. 
Further, the wall effects also result in a correction to the drag at \textit{O}($\alpha^2/H$) along the $z$-direction, leading to a point force at the same order in the equation. The point torque is zero till the order we have considered in our calculation. The force in (\ref{eq:wall force gravity aligned}) is presented till $O(1/H^3)$ assuming $\alpha=O(1/H)$. The assumption simplifies the analysis as the calculation becomes cumbersome if this scaling is not taken into account. For instance, the leading-order lift comes at $O(\alpha/H^2)$. If $\alpha\gg O(1/H)$, and say $\alpha^3= O(1/H)$, then to ensure consistency with the lift at leading-order, the calculation has to be carried out to include the corrections at $O(\alpha^4/H)$ as well. For this, one needs to use the modified boundary conditions at the sphere surface (section \ref{subsec:domain perturbation method}) at O($\alpha^4$) as well. Implementing the boundary conditions and performing the resulting calculations is cumbersome. Therefore, we assume $\alpha = O(1/H)$ to simplify the analysis.

Below, we discuss the scaling behaviour of the forces at different orders in $\alpha/H$. In the unbounded domain, the drag force arising from particle deformation scales [at $O(\alpha^2)$] as $(\mu^3 V_0^3/G^2)(1/R_0)$. Due to the wall effects, the drag force due to the deformation at the leading order scales [at $O(\alpha^2/H)$] as $(\mu^3 V_0^3/G^2)(1/h)$. The leading order lift appears at $O(\alpha/H^2)$, which scales as $(\mu^2 V_0^2/G)(R_0/h)^2$. The scaling for the lift can be seen as arising from the stresslet velocity field (that decays as $1/\xi^2$, as seen in (\ref{eq:vr1 unbounded})). One can estimate the scaling of the stresslet singularity ($S_t$) as $(\mu V_0 R_0)\times(\mu V_0/G)$, the term in the first bracket being the viscous force scaling, and that in the second is the scaling for the elastic deformation. The velocity field after reflecting from the wall scales as $S_t/(\mu h^2)$ and acts as an ambient for the elastic particle, which results in the force scaling given by $\mu [S_t/(\mu h^2)] R_0= (\mu^2 V_0^2/G)(R_0/h)^2$. For an elastic sphere moving parallel to a rigid wall in the lubrication limit, the scaling in the lubrication limit is $(\mu^2 V_0^2/G)(R_0/h)^{5/2}$ (expressed using the variables adopted in our analysis) \citep{skotheim2005soft}. Note that the decay rate of the lift force obtained in our calculation is different from that predicted in the lubrication limit.

\begin{figure}
  \centerline{\includegraphics[width=0.5\textwidth]{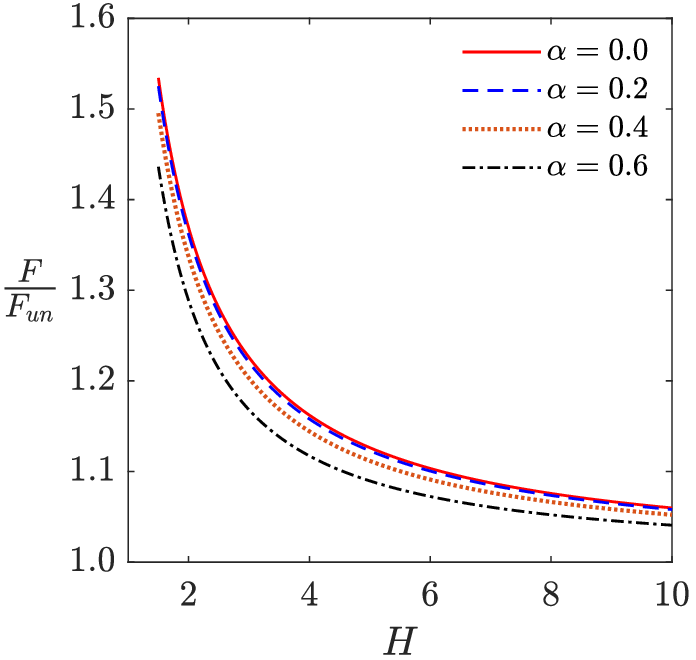}}
  \caption{Variation of external point force (parallel to the wall) given in (\ref{eq:wall force gravity aligned}) relative to the unbounded given in (\ref{eq:unbound force gravity aligned}) for different wall-particle gap widths, H varying from 1.5 to 10 at $\tilde{\rho}=1.05$, $K_{fz}=0.5$ and $\gamma=2$.}
\label{fig:drag variation with alpha and H}
\end{figure}

\begin{figure}
  \centerline{\includegraphics[width=1\textwidth]{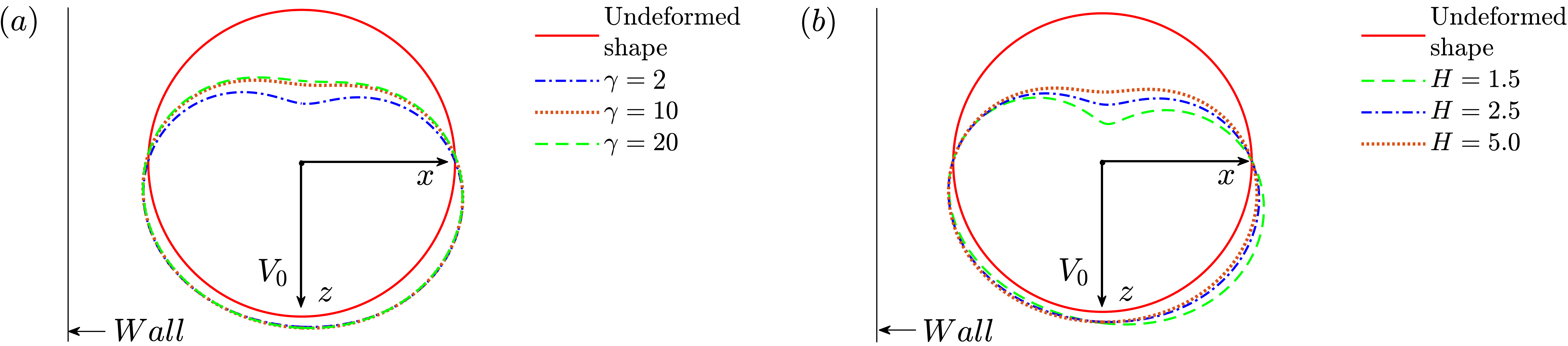}}
  \caption{Variation in the shape of the deformed sphere on $xz$ plane at $\alpha=0.2$; (\textit{a}) for different values of $\gamma$ at $H=2.5$  and (\textit{b}) for different values of $H$ at $\gamma=2$.}
\label{fig:def_bound_lbyG_H}
\end{figure}

Figure  \ref{fig:drag variation with alpha and H} shows the variation of $F/F_{un}$, ratio of point forces in the semi-infinite and unbounded domain acting along $z$-axis, with $H$, for different values of $\alpha$. The figure illustrates that the particle experiences greater resistance to the flow when it is near the wall than when it is away. Additionally, as the particle deformability increases, the force ratio decreases at a given $H$. 
The deformed shape of the particle at different values of $\gamma$ for fixed $H$ value at $\alpha=0.2$ is shown in figure \ref{fig:def_bound_lbyG_H}(a). As expected, particles with larger $\gamma$ exhibit reduced volumetric deformation compared to those with lower $\gamma$. Figure \ref{fig:def_bound_lbyG_H}(b) illustrates the variation of deformed shape with $H$ at a constant $\gamma$. As $H$ decreases, the particle experiences greater flow resistance, and the asymmetry in the deformed shape increases.

The point force for a neutrally buoyant particle can be obtained by equating the particle and fluid densities at the leading-order [$\tilde{\rho}=1$, defined below (\ref{eq:c2 expression gravity aligned})] in (\ref{eq:wall force gravity aligned}) and is obtained as
\begin{align}
     \mathbf{F}_{nb}&=6\pi\mu V_0 R_{0}\left[\left(1+\frac{9 }{16 H} +\frac{81}{256 H^2}+\frac{217}{4096 H^3}\right)\mathbf{\hat{k}}-\alpha\left(\frac{9}{64 H^2}\right) \mathbf{\hat{i}}+ \alpha^2\left(E_1\left[1+\frac{9}{4H}\right]+\right.\right. \nonumber\\& \left.\left. K_{fz}^3\, E_2 + K_{fz}^2 \left(E_3-\frac{27 (8 + 69 \gamma + 10 \gamma^2)}{160 (2 + \gamma)\,H}\right)E_2+K_{fz}\left[E_4E_2+\frac{E_5}{H}\right]\right)\mathbf{\hat{k}}\right] .\label{eq:wall force gravity aligned_neutrally buoyant}
\end{align}
Here, 
\begin{align}
E_5=-\frac{9 (1792 + 6354 \gamma + 7213 \gamma^2 + 2922 \gamma^3 + 475 \gamma^4)}{64(2 + \gamma) (2 + 3 \gamma)^2 (14 + 19 \gamma)}.
\end{align}
It is worth noting that even a neutrally buoyant particle experiences a buoyancy force at \textit{O}($\alpha^2$) due to its compressibility. In the absence of gravity or when $K_{fz}\ll1$, a limit relevant for microfluidic devices, (\ref{eq:wall force gravity aligned}) simplified further and one gets the force as
\begin{align}
    \mathbf{F}_{0g}= 6\pi\mu V_0 R_{0}\left[\left(1+\frac{9 }{16 H} +\frac{81}{256 H^2}+\frac{217}{4096 H^3}\right)\mathbf{\hat{k}}-\alpha\left(\frac{9}{64 H^2}\right) \mathbf{\hat{i}}+\alpha^2E_1\left(1+\frac{9}{4H}\right)\mathbf{\hat{k}} \right].\label{eq:wall force_0 gravity}
\end{align}

\subsection{Gravity in an arbitrary direction}\label{subsec:grav_arb}
Here, we present the point force and the point torque for the case of the particle translating parallel to the wall and gravity is in an arbitrary direction ($\mathbf{g}=g_x\mathbf{\hat{i}}+g_y\mathbf{\hat{j}}+g_z\mathbf{\hat{k}}$). Following the procedure outlined in section \ref{subsec:solution procedure}, we first calculated the variables in an unbounded domain and then accounted for wall effects using the method of reflections. Due to the misalignment between gravity and the particle velocity, the deformed shape of the particle lacks symmetry about the $z$-axis even in the unbounded domain. The point forces in unbounded and semi-infinite domains for arbitrary gravity, are represented by 
$\mathbf{F}_{un,ag}$ and $\mathbf{F}_{ag}$, respectively. The force in the unbounded domain is obtained as 
\begin{align}
    \mathbf{F}_{un,ag}&= 6\pi\mu V_0 R_{0}\left(\ \left[\left(1+\frac{2K_{fz}}{9}\left[1-\tilde{\rho} \right]\right)\mathbf{\hat{k}}+\frac{2[1-\tilde{\rho}]}{9}\left(K_{fx}\mathbf{\mathbf{\hat{i}}}+K_{fy}\mathbf{\hat{j}}\right)\right]+\right. \nonumber\\& \left.\alpha^2\left[F^{(2)}_{un,ag_x}\mathbf{\hat{i}}+F^{(2)}_{un,ag_y}\mathbf{\hat{j}}+F^{(2)}_{un,ag_z}\mathbf{\hat{k}}\right]\ \right). \label{eq:unbound force arbitrary gravity}
\end{align}
Above, the Cartesian components of the force coefficient at \textit{O}($\alpha^2$), denoted as $F^{(2)}_{un,ag_i}$, ($i=x,y,z$), are given in the Appendix \ref{app:gravity arbitrary}. Note that the force at \textit{O}($\alpha^2$) along $z$-direction is a function of $K_{fx}$, $K_{fy}$, $K_{fz}$, $\gamma$, and $\tilde{\rho}$ and is not the same as that at \textit{O}($\alpha^2$) in (\ref{eq:unbound force gravity aligned}), where it is only a function of $K_{fz}$, $\gamma$, and $\tilde{\rho}$. Therefore, the body forces in $x$- and $y$-directions also affect the force in the $z$-direction in the arbitrary gravity case, due to the non-linearity in the boundary conditions. The drag at $O(\alpha^2)$ scales as $(\mu^3 V_0^3/G^2)(1/R_0)$. Further, the torque comes at \textit{O}($\alpha$) and is obtained as
\begin{align}
    \mathbf{T}_{un,ag}= 8\pi\mu V_0 R_{0}^2\left(\alpha\left[\frac{2-2\gamma(1-\tilde{\rho})+\tilde{\rho}}{8(2+3\gamma)} \left(-K_{fy} \mathbf{\mathbf{\hat{i}}} +K_{fx} \mathbf{\hat{j}}\right)\right]\ \right). \label{eq:unbound torque arbitrary gravity}
\end{align}
Due to the elastic effects, the torque appears at \textit{O}($\alpha$) in this case, while it is zero when gravity is aligned with the particle velocity. The torque at \textit{O}($\alpha$) scales as $(\mu^2 V_0^2/G)(R_0)$.

The simplified form of the point force ($\mathbf{F}^x_{un,ag}$) and the point torque ($\mathbf{T}^x_{un,ag}$) when gravity acts along the $x$-axis ($\mathbf{g}=g_x\mathbf{\hat{i}}$) is obtained by substituting $K_{fy}=K_{fz}=0$ in (\ref{eq:unbound force arbitrary gravity}) and (\ref{eq:unbound torque arbitrary gravity}), respectively. The point force is obtained as
\begin{align}
    \mathbf{F}^x_{un,ag}&= 6\pi\mu V_0 R_{0}\left(\ \left[1\, \mathbf{\hat{k}}+\frac{2(1-\tilde{\rho})}{9}K_{fx}\mathbf{\mathbf{\hat{i}}}\right]+\alpha^2\left[F^{(2),x}_{un,ag_x}\mathbf{\hat{i}}+F^{(2),x}_{un,ag_z}\mathbf{\hat{k}}\right]\ \right). \label{eq:unbound force gravity only in x direction}
\end{align}
Here,
\begin{align}
    &F^{(2),x}_{un,ag_x}=\frac{-1}{144(2+\gamma)(2+3\gamma)^2(14+19\gamma)}\left(8K_{fx}^3[14+19\gamma]J_1+9K_{fx}J_2\right), \nonumber\\
    &F^{(2),x}_{un,ag_z}=E_1+K_{fx}^2 J_3,\nonumber\\
    &J_1= 18 + \gamma (45 + \gamma (16 + \gamma)) - 32 \tilde{\rho} - 2 \gamma (33 + \gamma (12 + \gamma)) \tilde{\rho} + (10 + \gamma (19 + \gamma (8 + \gamma))) \tilde{\rho}^2,\nonumber\\
    &J_2=28 (83 + 30 \tilde{\rho}) + \gamma (7008 + 3672 \tilde{\rho} + \gamma (6551 + 5658 \tilde{\rho} + \gamma (2010 + 287 \gamma + 3504 \tilde{\rho} + 663 \gamma \tilde{\rho}))),\nonumber\\
    &J_3=\frac{1}{40(2+\gamma)(2+3\gamma)^2(14+19\gamma)}[-217 \gamma^4 (-1 + \tilde{\rho})^2 - 2 \gamma^3 (-1 + \tilde{\rho}) (-891 + 665 \tilde{\rho}) + \nonumber\\& \gamma^2 (-4609 + (7564 - 2803 \tilde{\rho}) \tilde{\rho}) - 28 (33 + \tilde{\rho} (-64 + 23 \tilde{\rho})) - 8 \gamma (471 + \tilde{\rho} (-811 + 288 \tilde{\rho}))],
\end{align}
and the point torque is obtained as
\begin{align}
    \mathbf{T}^x_{un,ag}= 8\pi\mu V_0 R_{0}^2\left(\alpha\left[\frac{2-2\gamma(1-\tilde{\rho})+\tilde{\rho}}{8(2+3\gamma)} K_{fx} \mathbf{\hat{j}}\right]\ \right). \label{eq:unbound torque gravity in x direction}
\end{align}
{Above, superscript $()^x$ denotes the gravity is in the $x$-direction. Due to the non-linearity in the boundary conditions, the point force at \textit{O}($\alpha^2$) in the $z$-direction depends on the gravitational force in the $x$-direction, as given in (\ref{eq:unbound force gravity only in x direction}). 

\begin{figure}
  \centerline{\includegraphics[width=1\textwidth]{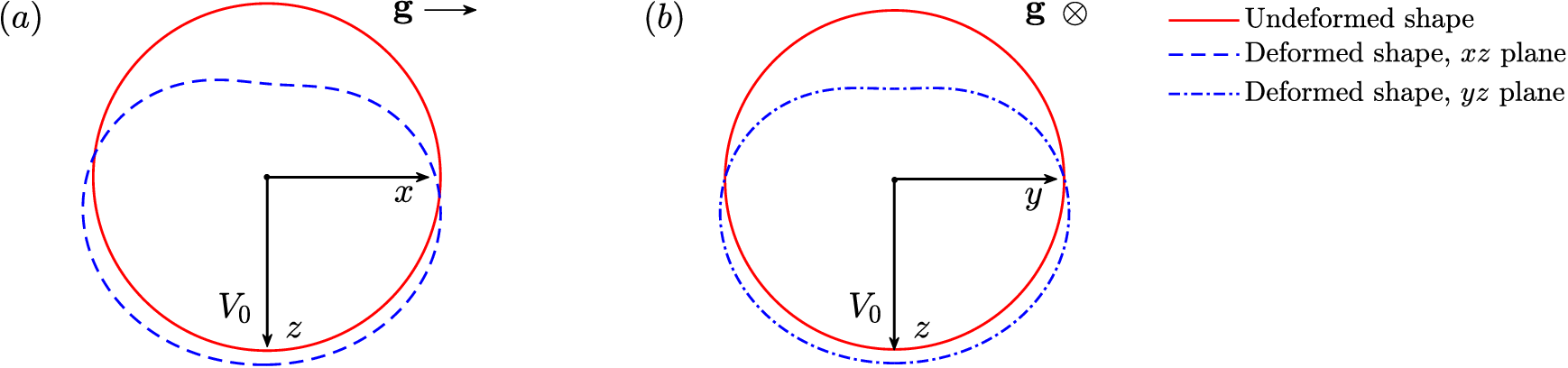}}
  \caption{Variation in the shape of the deformed sphere in the unbounded domain at $K_{fx}=2$, $\alpha=0.2$, $\gamma=2$ and $\tilde{\rho}=1.05$ on (\textit{a}) $xz$ plane and (\textit{b}) $yz$ plane.}
\label{fig:def_bound_Kfx_xz_yz plane}
\end{figure}

To illustrate the effect of gravitational force orientation on particle deformation, we plotted the deformed shape in an unbounded domain on the $xz$ and $yz$ planes when gravity is perpendicular to the particle velocity and is acting in the $x$-direction. When gravity acts along the $x$-axis, the shape on the $xz$ and $yz$ planes is shown in figure \ref{fig:def_bound_Kfx_xz_yz plane}(a) and (b), respectively. {On the plane $y=0$, the deformed shape remains symmetric about the $z$-axis in the absence of gravity. However, as shown in figure \ref{fig:def_bound_Kfx_xz_yz plane}(a), when gravity acts in the $x$-direction, perpendicular to the velocity of the particle, the deformed shape becomes asymmetric about the $z$-axis. This causes a shift in the centroid of the particle relative to that of the undeformed shape, generating a fluid-induced moment on the particle. This moment is balanced by the external point torque, to make the particle torque-free.} However, on the plane $x=0$, the deformed shape remains symmetric about the $z$-axis, consistent with the symmetry of the surrounding flow field.

We then analysed the wall effects on the particle dynamics for the arbitrary gravity case. In the presence of the wall, the velocity and displacement fields are calculated at orders $1/H$, $\alpha/H$, and $1/H^2$. The point force till the order we considered is obtained as
\begin{align}
    \mathbf{F}_{ag}&= \mathbf{F}_{un,ag}+6\pi\mu V_0 R_{0}\left[\left(\frac{9 }{16 H} +\frac{81}{256 H^2}\right)\mathbf{\hat{k}}\right].\label{eq:bounded force arbitrary gravity}
\end{align}
The force at \textit{O}($\alpha/H$) is 0 and the point torque is obtained as
\begin{align}
   \mathbf{T}_{ag}=\mathbf{T}_{un,ag}\left[1+\frac{9}{16H}\right].\label{eq:bounded torque arbitrary gravity}
\end{align}
The torque at \textit{O}($\alpha/H$) scales as $(\mu^2 V_0^2/G)/(R_0^2/h)$. The first effects of the wall appear as a torque at \textit{O}($\alpha/H$) in this case, as opposed to a lift at \textit{O}($\alpha/H^2$) in the case where gravity is aligned with the particle velocity (\ref{eq:wall force gravity aligned}).

\section{Summary}\label{sec:summary}
In this paper, we have investigated the dynamics of a weakly elastic spherical particle translating parallel to a rigid wall in a quiescent fluid in the limit, $\alpha \ll 1$. The motion of the particle is constrained by applying an external point force and point torque. Our model is useful to understand the dynamics of elastic beads embedded with magnetic particles. The analysis is presented for two cases: first, when the gravity is aligned with particle velocity and second, when the gravity is in an arbitrary direction.\\

For the first case, we have considered three reflections from the wall. The velocity and displacement fields calculated at orders $1/H$, $\alpha/H$, $\alpha^2/H$, $1/H^2$, $\alpha/H^2$, and $1/H^3$ are presented in section \ref{subsec:grav_parallel}. The external point force needed to make the particle translate in an unbounded domain and parallel to the wall are given in (\ref{eq:unbound force gravity aligned}) and (\ref{eq:wall force gravity aligned}), respectively. In the unbounded domain, the effects of elasticity appear as a drag at \textit{O}($\alpha^2$), and the external force is aligned with gravity ($z$-axis). The external point torque is zero till the order we considered in our calculation. A hydrodynamic lift arises due to the wall effects (away from the wall) at \textit{O}($\alpha/H^2$). The scaling of the lift force arises due to the stresslet singularity resulting from the deformation as discussed in section \ref{subsec:grav_parallel}. The decay rate for the lift force in our calculation scales as $(\mu^2 V_0^2/G)(R_0/h)^2$ and is different from that predicted in the lubrication limit, which scales as $(\mu^2 V_0^2/G)(R_0/h)^{5/2}$ by \citep{skotheim2005soft}. We note here that the lateral migration of the deformable particles, such as drops, vesicles or capsules, in bounded simple shear flow is the consequence of the stresslet singularity \citep{nix2014lateral,bureau2023lift}.
For a neutrally buoyant particle, the force expression (\ref{eq:wall force gravity aligned_neutrally buoyant}) highlights the emergence of buoyancy force at  \textit{O}($\alpha^2$) due to the compressibility of the particle. The expression in the limit of $K_{fz} \ll 1$ is given in (\ref{eq:wall force_0 gravity}). The force provided in the equation is particularly relevant for microfluidic devices, where one expects $K_{fz} \ll 1$. Figure \ref{fig:def_unbound_bound} shows the deformed shape of the particle in the unbounded domain and in the presence of the wall. The deformed shape is symmetric about the translation velocity ($z$-axis) in the unbounded domain, whereas it is asymmetric in the presence of the wall. The particle shape depends on the density ratio ($\tilde{\rho}$). For a heavier (lighter) particle, increasing $K_{fz}$ reduces (increases) the required external force and decreases (increases) the dip in the particle's upper section seen in figure \ref{fig:gravity strength variatio unbounded}.\\

For the second case, when gravity and particle velocity are not aligned, we considered two reflections from the wall. The velocity and displacement fields are calculated at orders $1/H$, $\alpha/H$, and $1/H^2$. 
The external force needed to translate the particle in an unbounded domain is given in (\ref{eq:unbound force arbitrary gravity}). The force simplifies to (\ref{eq:unbound force gravity only in x direction}) when gravity acts only in the $x$-direction. It is important to note that due to non-linear boundary conditions, the force at \textit{O}($\alpha^2$) in (\ref{eq:unbound force gravity only in x direction}) acts in the $z$-direction and depends on the gravitational force in the $x$-direction.
The external force and torque needed to translate the particle parallel to the wall are given in (\ref{eq:bounded force arbitrary gravity}) and (\ref{eq:bounded torque arbitrary gravity}), respectively. It is noted that for this case, the torque given in (\ref{eq:unbound torque arbitrary gravity}) comes at \textit{O}($\alpha$) in an unbounded domain, which is absent in the first case. Further, the elastic induced wall effects appear as torque at \textit{O}($\alpha/H$) as given in (\ref{eq:bounded torque arbitrary gravity}) for this case, in contrast to lift at \textit{O}($\alpha/H^2$) given in (\ref{eq:wall force gravity aligned}) for the former case. The analysis we have presented for an arbitrary gravity case can also be utilised for other body forces such as electric and magnetic forces, by replacing the gravitational force density with an electric or magnetic force density in the definition of $K_{px}$, $K_{py}$ and $K_{pz}$, defined below (\ref{eq:uphi displacement series}).

The linearity of Stokes flow allows one to derive a relationship between the hydrodynamic forces and torques on a particle and its translational and angular velocities through resistance and mobility formulations. The resistance problem, fundamental for determining hydrodynamic forces and torques with prescribed particle velocity, is particularly relevant in scenarios with imposed boundary conditions. Conversely, the mobility problem determines particle motion from applied forces and torques. These formulations are inverse to each other when the particle shape is the same. In both the resistance and mobility problems, the particle is in a state of dynamic equilibrium. To solve the problems, a knowledge of the force distribution that leads to the equilibrium is not needed for a rigid particle; however, in the case of an elastic particle, the knowledge becomes crucial. Notably, the deformed shape of an elastic particle depends on the distribution. The resistance problem of a weakly deformable particle moving parallel to a wall has been solved for both unbounded and semi-infinite domains in this work. In the unbounded case, our results differ from Murata \citep{murata1980deformation}, which employs a mobility formulation. While Murata predicts a prolate spheroid shape, our particle shape, shown in figure \ref{fig:def_unbound_bound}(a), is distinct. The difference arises from the varying force distributions. Further, our approach considers the effect of particle compressibility on density, which is not considered in Murata’s analysis when calculating the terminal velocity of the particle. We have revisited Murata's calculation and calculated the terminal velocity considering the density change of the particle, and the results are presented in the Appendix \ref{app:modification in Murata analysis}. 

Our findings highlight the importance of force distribution in determining the steady-state morphology of deformable particles moving in a fluid. Therefore, identifying the force distribution acting on a deformable particle is essential for analysing its dynamics. The theoretical framework developed in this paper using the point force and the point torque model can be extended to understand the dynamics of magnetoresponsive polymer beads in channels. While in the current work, we model such beads as linearly elastic, such an extension needs the use of non-linear elastic/viscoelastic models for the bead. The extension will allow one to estimate the magnetic force/torque needed to manipulate the beads in a microchannel.
\appendix
\section{Solutions to the Navier elasticity equations for a point force and a point torque}\label{app:displacement field for point force and point torque}
The governing equation for an infinite linear, isotropic and homogeneous elastic medium acted upon by a point force at $\mathbf{x=0}$ is given by
\begin{align}
  \mathbf[\lambda+G]  \mathbf{\nabla}\left( \mathbf{\nabla} \cdot  \mathbf{u}^{p_f}\right)+G  \mathbf{\nabla}^2  \mathbf{u}^{p_f}+\mathbf{F}\delta(\mathbf{x}) ={0}.\label{eq:elasticity with just point force}
\end{align}
Above, the displacement field ($\mathbf{u}^{p_f}$) is due to the point force and is given by
\begin{align}
    \mathbf{u}^{p_f}(\mathbf{x})&=\boldsymbol{G}(\mathbf{x})\cdot\mathbf{F}\nonumber \\
    &=\frac{1}{16(1-\nu)\pi G}\left[\frac{\mathbf{I}(3-4\nu)}{r}+\frac{\mathbf{x}\mathbf{x}}{r^3}\right]\cdot\mathbf{F} \, .\label{eq:green}
\end{align}
Here, $r=|\mathbf{x}|$, $\boldsymbol{G}(\mathbf{x})$ is the Green's function and $\nu=\lambda/(2[\lambda+G])$ is the Poisson ratio. It can be shown that (\ref{eq:green})
is indeed the solution by substituting it in (\ref{eq:elasticity with just point force}) and integrating  over any volume ($\mathring{V}$) enclosing point force. Using the divergence theorem, the integration yields 
\begin{align}
    \int_{\mathring{V}} \left(\mathbf[\lambda+G]  \mathbf{\nabla}\left( \mathbf{\nabla} \cdot  \mathbf{u}^{p_f}\right)+G  \mathbf{\nabla}^2  \mathbf{u}^{p_f}\right)dV=-\mathbf{F}\,.
\end{align}
One can derive the displacement field corresponding to a point torque by drawing an analogy to the rotlet singularity in the case of the Stokes equation \citep{kim2013microhydrodynamics}.
The governing equation for an infinite elastic medium with point torque at $\mathbf{x=0}$ is given by  
\begin{align}
  \mathbf[\lambda+G]  \mathbf{\nabla}\left( \mathbf{\nabla} \cdot  \mathbf{u}^{p_t}\right)+G  \mathbf{\nabla}^2  \mathbf{u}^{p_t}+\frac{1}{2}\left[\mathbf{\nabla}\delta(\mathbf{x})\times \mathbf{T} \right] ={0}\,.\label{eq:elasticity with just point torque}
\end{align}
Above, the displacement field ($\mathbf{u}^{p_t}$) is due to the point torque and is given by
\begin{align}
    \mathbf{u}^{p_t}(\mathbf{x})=\frac{\mathbf{T}\times\mathbf{x}}{8\pi G r^3}\,.\label{eq:torquegreen}
\end{align}
One can verify that the displacement field is indeed correct by substituting (\ref{eq:torquegreen}) in (\ref{eq:elasticity with just point torque}) and taking its moment about the origin and integrating over an arbitrary volume ($\mathring{V}$) enclosing the point torque. Using the divergence theorem, the integration yields 
\begin{align}
    \int_{\mathring{V}} \mathbf{x} \times \left(\mathbf[\lambda+G]  \mathbf{\nabla}\left( \mathbf{\nabla} \cdot  \mathbf{u}^{p_t}\right)+G  \mathbf{\nabla}^2  \mathbf{u}^{p_t}\right)dV=-\mathbf{T}\, .
\end{align} 

The displacement field corresponding to the point force and the point torque decays as $r^{-1}$ and $r^{-2}$, respectively.
\section{Gravity aligned with particle velocity }\label{app:gravity aligned}
The radial ($v^{(2)}_{un,r}$) and tangential ($v^{(2)}_{un,\theta}$) velocity components at \textit{O}($\alpha^2$) for the case when gravity is in $z$-direction are obtained as
\begin{align}
   v^{(2)}_{un,r}&=-\frac{3 \cos\theta}{320 (2 + \gamma)^2 (2 + 3 \gamma)^2 (14 + 19 \gamma) \xi^5} [- ((-1 + \xi^2) ( 45 (-1120 \xi^2 + \gamma (368 - 3440 \xi^2) +\nonumber\\& \gamma^3 (1881 - 860 \xi^2) + 8 \gamma^4 (131 + 4 \xi^2) + \gamma^5 (177 + 4 \xi^2) - 22 \gamma^2 (-63 + 148 \xi^2)) - 60 K_{fz} (\nonumber\\&3 \gamma^5 (23 + 28 \xi^2) (-1 + \tilde{\rho}) - 112 (2 + \xi^2 (-3 + 2 \tilde{\rho})) + 6 \gamma^2 (-197 + 63 \tilde{\rho} + 6 \xi^2 (17 + 6 \tilde{\rho})) +\nonumber\\& \gamma (-824 + 92 \tilde{\rho} - 64 \xi^2 (-16 + 7 \tilde{\rho})) + \gamma^4 (-400 + 358 \tilde{\rho} + 8 \xi^2 (-59 + 65 \tilde{\rho})) + \gamma^3 (-901 +\nonumber\\& 570 \tilde{\rho} + \xi^2 (-516 + 896 \tilde{\rho}))) + 4 K_{fz}^2 (\gamma^5 (81 + 244 \xi^2) (-1 + \tilde{\rho})^2 + 8 \gamma^4 (-1 + \tilde{\rho}) (-66 + 39 \tilde{\rho} +\nonumber\\& \xi^2 (-109 + 186 \tilde{\rho})) - 4 \gamma (-340 + 586 \tilde{\rho} + 51 \tilde{\rho}^2 + 4 \xi^2 (315 - 349 \tilde{\rho} + \tilde{\rho}^2)) - 56 (-4 + 12 \tilde{\rho} +\nonumber\\& \tilde{\rho}^2 + 4 \xi^2 (6 - 8 \tilde{\rho} + \tilde{\rho}^2)) + 2 \gamma^2 (1149 - 1474 \tilde{\rho} - 80 \tilde{\rho}^2 + 36 \xi^2 (-74 + 54 \tilde{\rho} + 25 \tilde{\rho}^2)) + \gamma^3 (\nonumber\\&3 (503 - 628 \tilde{\rho} + 68 \tilde{\rho}^2) + 4 \xi^2 (-201 - 499 \tilde{\rho} + 719 \tilde{\rho}^2))))) + 5 ( 9 (2800 \xi^2 + 40 \gamma (-46 + 365 \xi^2)\nonumber\\& + 8 \gamma^4 (-655 + 1319 \xi^2) + \gamma^5 (-885 + 1493 \xi^2) + \gamma^3 (-9405 + 25757 \xi^2) + \gamma^2 (-6930 +\nonumber\\& 28498 \xi^2)) - 12 K_{fz} (\gamma^5 (-345 + 649 \xi^2) (-1 + \tilde{\rho}) + 560 (2 + \xi^2 (-3 + 2 \tilde{\rho})) + 2 \gamma^4 (1000 - \nonumber\\&895 \tilde{\rho} + \xi^2 (-2176 + 2147 \tilde{\rho})) + 2 \gamma^2 (2955 - 945 \tilde{\rho} + \xi^2 (-6439 + 5563 \tilde{\rho})) + \gamma (4120 - 460 \tilde{\rho} +\nonumber\\& \xi^2 (-7456 + 5732 \tilde{\rho})) + \gamma^3 (4505 - 2850 \tilde{\rho} + \xi^2 (-10785 + 10154 \tilde{\rho}))) + 4 K_{fz}^2 (\gamma^5 (-81 +\nonumber\\& 233 \xi^2) (-1 + \tilde{\rho})^2 + 8 \gamma^4 (-1 + \tilde{\rho}) (66 - 39 \tilde{\rho} + 2 \xi^2 (-97 + 93 \tilde{\rho})) + 56 (-4 + 12 \tilde{\rho} + \tilde{\rho}^2 + \xi^2 (\nonumber\\&6 - 20 \tilde{\rho} + 7 \tilde{\rho}^2)) + 4 \gamma (-340 + 586 \tilde{\rho} + 51 \tilde{\rho}^2 + \xi^2 (546 - 1270 \tilde{\rho} + 493 \tilde{\rho}^2)) + 2 \gamma^2 (-1149 +\nonumber\\& 1474 \tilde{\rho} + 80 \tilde{\rho}^2 + \xi^2(2221 - 4430 \tilde{\rho} + 1894 \tilde{\rho}^2))+ \gamma^3 (-3 (503 - 628 \tilde{\rho} + 68 \tilde{\rho}^2) + \xi^2 (3853 -\nonumber\\& 7444 \tilde{\rho} + 3458 \tilde{\rho}^2)))) \cos(2\theta) ], \label{eq:app v2r unbounded}
\end{align}
and
\begin{align}
    v^{(2)}_{un,\theta}&=-\frac{3}{1280 (2 + \gamma)^2 (2 + 3 \gamma)^2 (14 + 19 \gamma) \xi^5} [ 45 ( 560 \xi^2 (1 - 4 \xi^2) + \gamma^3 (-16929 + 20275 \xi^2 -\nonumber\\& 1720 \xi^4) + 8 \gamma^4 (-1179 + 1065 \xi^2 + 8 \xi^4) + \gamma^5 (-1593 + 1147 \xi^2 + 8 \xi^4) - 8 \gamma (414 - 873 \xi^2 +\nonumber\\& 860 \xi^4) - 2 \gamma^2 (6237 - 9607 \xi^2 + 3256 \xi^4) ) - 60 K_{fz} ( \gamma^5 (-621 + 679 \xi^2 + 168 \xi^4) (-1 + \tilde{\rho}) - \nonumber\\&112 (-18 + \xi^2 (5 - 6 \tilde{\rho}) + \xi^4 (-6 + 4 \tilde{\rho})) - 4 \gamma (\xi^2 (940 - 1163 \tilde{\rho}) + 32 \xi^4 (-16 + 7 \tilde{\rho}) + 9 (-206\nonumber\\& + 23 \tilde{\rho})) + 2 \gamma^4 (1800 - 1611 \tilde{\rho} + 8 \xi^4 (-59 + 65 \tilde{\rho}) + \xi^2 (-2248 + 2309 \tilde{\rho})) + 2 \gamma^2 (5319 - \nonumber\\&1701 \tilde{\rho} + 36 \xi^4 (17 + 6 \tilde{\rho}) + \xi^2 (-4645 + 5401 \tilde{\rho})) + \gamma^3 (8109 - 5130 \tilde{\rho} + 8 \xi^4 (-129 + 224 \tilde{\rho}) + \nonumber\\&\xi^2 (-10015 + 10806 \tilde{\rho})) ) + 4 K_{fz}^2 ( \gamma^5 (-729 + 1491 \xi^2 + 488 \xi^4) (-1 + \tilde{\rho})^2 + 8 \gamma^4 (-1 + \tilde{\rho}) \nonumber\\&(594 - 351 \tilde{\rho} + 24 \xi^2 (-44 + 51 \tilde{\rho}) + \xi^4 (-218 + 372 \tilde{\rho})) - 56 (\xi^2 (26 + 12 \tilde{\rho} - 29 \tilde{\rho}^2) + 8 \xi^4 (\nonumber\\&6 - 8 \tilde{\rho} + \tilde{\rho}^2) - 9 (-4 + 12 \tilde{\rho} + \tilde{\rho}^2)) - 4 \gamma (\xi^2 (470 + 2386 \tilde{\rho} - 2559 \tilde{\rho}^2) + 8 \xi^4 (315 - 349 \tilde{\rho} +\nonumber\\& \tilde{\rho}^2) - 9 (-340 + 586 \tilde{\rho} + 51 \tilde{\rho}^2)) + 2 \gamma^2 (72 \xi^4 (-74 + 54 \tilde{\rho} + 25 \tilde{\rho}^2) + 9 (-1149 + 1474 \tilde{\rho} +\nonumber\\& 80 \tilde{\rho}^2) + \xi^2 (3479 - 15314 \tilde{\rho} + 11430 \tilde{\rho}^2)) + \gamma^3 (-27 (503 - 628 \tilde{\rho} + 68 \tilde{\rho}^2) + 8 \xi^4 (-201 - 499 \tilde{\rho}\nonumber\\& + 719 \tilde{\rho}^2) + \xi^2 (14639 - 37444 \tilde{\rho} + 22634 \tilde{\rho}^2)) ) + 5 ( 9 (2800 \xi^2 + 40 \gamma (-138 + 365 \xi^2) + 8 \gamma^4 (\nonumber\\&-1965 + 1319 \xi^2) + \gamma^5 (-2655 + 1493 \xi^2) + \gamma^3 (-28215 + 25757 \xi^2) + \gamma^2 (-20790 + \nonumber\\&28498 \xi^2)) - 12 K_{fz} (\gamma^5 (-1035 + 649 \xi^2) (-1 + \tilde{\rho}) + 560 (6 + \xi^2 (-3 + 2 \tilde{\rho})) + 4 \gamma (3090 -\nonumber\\& 345 \tilde{\rho} + \xi^2 (-1864 + 1433 \tilde{\rho})) + \gamma^4 (6000 - 5370 \tilde{\rho} + \xi^2 (-4352 + 4294 \tilde{\rho})) + 2 \gamma^2 (8865 - \nonumber\\&2835 \tilde{\rho} + \xi^2 (-6439 + 5563 \tilde{\rho})) + \gamma^3 (13515 - 8550 \tilde{\rho} + \xi^2 (-10785 + 10154 \tilde{\rho}))) + 4 K_{fz}^2 (\nonumber\\&\gamma^5 (-243 + 233 \xi^2) (-1 + \tilde{\rho})^2 + 8 \gamma^4 (-1 + \tilde{\rho}) (198 - 117 \tilde{\rho} + 2 \xi^2 (-97 + 93 \tilde{\rho})) + 56 (3 (-4 +\nonumber\\& 12 \tilde{\rho} + \tilde{\rho}^2) + \xi^2 (6 - 20 \tilde{\rho} + 7 \tilde{\rho}^2)) + 4 \gamma (3 (-340 + 586 \tilde{\rho} + 51 \tilde{\rho}^2) + \xi^2 (546 - 1270 \tilde{\rho} + 493 \tilde{\rho}^2)) \nonumber\\&+ \gamma^3 (-9 (503 - 628 \tilde{\rho} + 68 \tilde{\rho}^2) + \xi^2 (3853 - 7444 \tilde{\rho} + 3458 \tilde{\rho}^2)) + \gamma^2 (-6894 + 8844 \tilde{\rho} +\nonumber\\& 480 \tilde{\rho}^2 + \xi^2 (4442 - 8860 \tilde{\rho} + 3788 \tilde{\rho}^2))) ) \cos(2\theta) ] \sin\theta. \label{eq:app v2theta unbounded}
\end{align}
The expressions for $E_1$, $E_2$, $E_3$, $E_4$, $I_1$, $I_2$, $I_3$, $I_4$ and $I_5$ in (\ref{eq:unbound force gravity aligned}) and (\ref{eq:wall force gravity aligned}) for both the unbounded and semi-infinite case are obtained as
\begin{align}
    &E_1=\frac{9  (-140 - 360 \gamma - 227 \gamma^2 + 6 \gamma^3 + \gamma^4)}{8 (2 + \gamma) (2 + 3 \gamma)^2 (14 + 19 \gamma)},\ E_2=\frac{28 + 52 \gamma + 19 \gamma^2}{9 (2 + \gamma) (2 + 3 \gamma)^2 (14 + 19 \gamma)},\label{appeq:E1 E2 expression}\\
    &E_3=-\frac{3 (7 + \gamma) (1 + 5 \gamma)}{5 (2 + \gamma)},\ E_4=-\frac{9 (1708 + 5964 \gamma + 6787 \gamma^2 + 2850 \gamma^3 + 475 \gamma^4)}{8 (28 + 52 \gamma + 19 \gamma^2)},\label{appeq:E3 E4 expression}\\
    &I_1=-\frac{\gamma^3 (1 - \tilde{\rho})^2 + 8 \gamma^2 (2 - 3 \tilde{\rho} + \tilde{\rho}^2) + 2 (9 - 16 \tilde{\rho} + 5 \tilde{\rho}^2) + \gamma (45 - 66 \tilde{\rho} + 19 \tilde{\rho}^2)}{2 (2 + \gamma)},\label{appeq:I1 expression}\\
    &I_2=\frac{1}{2 (7 + \gamma) (1 + 5 \gamma) (14 + 19 \gamma)}\left[\gamma^3 (3102 - 2162 \tilde{\rho} - 750 \tilde{\rho}^2) + \gamma^2 (10734 - 8569 \tilde{\rho} - \right. \nonumber\\& \left.657 \tilde{\rho}^2) + \gamma^4 (292 - 109 \tilde{\rho} - 183 \tilde{\rho}^2) + 28 (108 - 104 \tilde{\rho} + 3 \tilde{\rho}^2) - 2 \gamma (-5224 + 4569 \tilde{\rho} + 18 \tilde{\rho}^2)\right],\label{appeq:I2 expression}\\
    &I_3=\frac{2044 - 336 \tilde{\rho} - 84 \gamma (-77 + 6 \tilde{\rho}) + 6 \gamma^3 (299 + 176 \tilde{\rho}) + \gamma^4 (223 + 252 \tilde{\rho}) + \gamma^2 (6211 + 576 \tilde{\rho})}{1708 + 5964 \gamma + 6787 \gamma^2 + 2850 \gamma^3 + 475 \gamma^4},\label{appeq:I3 expression}\\
    &I_4=\frac{27}{160 (28 + 52 \gamma + 19 \gamma^2)} \left( \gamma^4 (-1 + \tilde{\rho}) (109 + 366 \tilde{\rho}) + 2 \gamma^3 (-1407 + 562 \tilde{\rho} + 750 \tilde{\rho}^2) - \right. \nonumber\\& \left.56 (63 + \tilde{\rho} (-64 + 3 \tilde{\rho})) + 2 \gamma (-6043 + 12 \tilde{\rho} (454 + 3 \tilde{\rho})) + \gamma^2 (-11913 + 2 \tilde{\rho} (4574 + 657 \tilde{\rho})) \right),\label{appeq:I4 expression}\\
    &I_5=-\frac{9}{64 (2 + \gamma) (2 + 3 \gamma)^2 (14 + 19 \gamma)} \left( \gamma (7110 - 756 \tilde{\rho}) + 56 (41 - 9 \tilde{\rho}) + 6 \gamma^3 (223 + 264 \tilde{\rho}) + \right. \nonumber\\& \left.\gamma^4 (97 + 378 \tilde{\rho}) + \gamma^2 (6349 + 864 \tilde{\rho}) \right).\label{appeq:I5 expression}
\end{align}
The expressions for $B_1$, $B_2$, $B_3$, $B_4$ and $B_5$ in (\ref{eq:wall induced deformation gravity aligned}) are obtained as
\begin{align}
    &B_1=\frac{27 (5 + 8 \gamma + 2 \gamma^2)}{32 (2 + \gamma) (2 + 3 \gamma)},\ B_2=-\frac{81 (7840 + 33504 \gamma + 53806 \gamma^2 + 40067 \gamma^3 + 13704 \gamma^4 + 1739 \gamma^5)}{512  (2 + \gamma)^2 (2 + 3 \gamma)^2 (14 + 19 \gamma)},\label{appeq:B1 B2 expression}\\
    &B_3=\frac{27}{256  (2 + \gamma)^2 (2 + 3 \gamma)^2 (14 + 19 \gamma)}\left(647 \gamma^5 (-1 + \tilde{\rho}) + 224 (-11 + 12 \tilde{\rho}) + 4 \gamma (-2798 + \right. \nonumber\\& \left.2929 \tilde{\rho}) + \gamma^4 (-4960 + 5042 \tilde{\rho}) + 2 \gamma^2 (-9353 + 9587 \tilde{\rho}) + \gamma^3 (-14231 + 14526 \tilde{\rho})\right),\label{appeq:B3 expression}\\
    &B_4=-\frac{81 (5600 + 27360 \gamma + 50066 \gamma^2 + 42109 \gamma^3 + 15864 \gamma^4 + 2101 \gamma^5)}{512  (2 + \gamma)^2 (2 + 3 \gamma)^2 (14 + 19 \gamma)},\label{appeq:B4 expression}\\
    &B_5=\frac{27}{256  (2 + \gamma)^2 (2 + 3 \gamma)^2 (14 + 19 \gamma)} \left(2240 (-1 + \tilde{\rho}) + 953 \gamma^5 (-1 + \tilde{\rho}) + 4 \gamma (-2698 +\right. \nonumber\\& \left. 2751 \tilde{\rho}) + \gamma^4 (-6704 + 6798 \tilde{\rho}) + 2 \gamma^2 (-9923 + 10181 \tilde{\rho}) + \gamma^3 (-17065 + 17458 \tilde{\rho})\right).\label{appeq:B5 expression}
\end{align}
\section{Gravity in an arbitrary direction}\label{app:gravity arbitrary}
The expression for the components of force coefficient at \textit{O}($\alpha^2$) given in (\ref{eq:unbound force arbitrary gravity}) where gravity is in the arbitrary direction ($\mathbf{g}=g_x\mathbf{\hat{i}}+g_y\mathbf{\hat{j}}+g_z\mathbf{\hat{k}}$) are obtained as
\begin{align}
F^{(2)}_{un,ag_x}&=\frac{K_{fx}}{720(2+\gamma)(2+3\gamma)^2 (14+19\gamma)} ( \gamma [ -45\gamma(6551+\gamma(2010+287\gamma)) - \nonumber\\&
40 K_{fz}^2(972+\gamma(1079+\gamma(318+19\gamma))) -6 K_{fz}(30488+\gamma(29109+ \nonumber\\&
11\gamma(642+47\gamma))) ] + ( 80 K_{fz}^2(14+19\gamma)(16+\gamma(33+\gamma(12+\gamma))) - \nonumber\\&
135\gamma(1224+\gamma(1886+\gamma(1168+221\gamma))) -12 K_{fz}(-3136+\gamma(-8544+ \nonumber\\&
\gamma(-5792+\gamma(344+433\gamma))))) \tilde{\rho} + 2 K_{fz} [ -20 K_{fz}(14+19\gamma)(10+\gamma(19+ \nonumber\\&
\gamma(8+\gamma))) + 9(532+\gamma(2352+\gamma(3679+\gamma(2330+461\gamma)))) ] \tilde{\rho}^2 - \nonumber\\&
36(2905+1554 K_{fz}+280 K_{fz}^2 +8760\gamma+1050\tilde{\rho}) -40 K_{fx}^2(14+19\gamma) ( 18+ \nonumber\\&
\gamma(45+\gamma(16+\gamma))-32\tilde{\rho}-2\gamma(33+\gamma(12+\gamma))\tilde{\rho} + (10+\gamma(19+\gamma(8+\gamma)))\tilde{\rho}^2 ) - \nonumber\\&
40 K_{fy}^2(14+19\gamma) ( 18+\gamma(45+\gamma(16+\gamma))-32\tilde{\rho}-2\gamma(33+\gamma(12+\gamma))\tilde{\rho} + \nonumber\\&
(10+\gamma(19+\gamma(8+\gamma)))\tilde{\rho}^2 ) ), \label{eq:F2un_x gravity arbitrary}
\end{align}
\begin{align}
     F^{(2)}_{un,ag_y}=\frac{K_{fy}}{K_{fx}}\left[F^{(2)}_{un,ag_x}\right],
     \label{eq:F2un_y gravity arbitrary}
\end{align}
\begin{align}
F^{(2)}_{un,ag_z} &=-\frac{1}{360 (2 + \gamma) (2 + 3 \gamma)^2 (14 + 19 \gamma)} ( 8316 K_{fy}^2 + 5040 K_{fy}^2 K_{fz} + 36288 K_{fz}^2 + 5040 K_{fz}^3 \nonumber\\&
+ 33912 K_{fy}^2 \gamma + 291060 K_{fz} \gamma + 19440 K_{fy}^2 K_{fz} \gamma + 125376 K_{fz}^2 \gamma + 19440 K_{fz}^3 \gamma + 91935 \gamma^2 \nonumber\\&
+ 41481 K_{fy}^2 \gamma^2 + 279495 K_{fz} \gamma^2 + 21580 K_{fy}^2 K_{fz} \gamma^2 + 128808 K_{fz}^2 \gamma^2 + 21580 K_{fz}^3 \gamma^2 \nonumber\\&
- 2430 \gamma^3 + 16038 K_{fy}^2 \gamma^3 + 80730 K_{fz} \gamma^3 + 6360 K_{fy}^2 K_{fz} \gamma^3 + 37224 K_{fz}^2 \gamma^3 + 6360 K_{fz}^3 \gamma^3 \nonumber\\&
- 405 \gamma^4 + 1953 K_{fy}^2 \gamma^4 + 10035 K_{fz} \gamma^4 + 380 K_{fy}^2 K_{fz} \gamma^4 + 3504 K_{fz}^2 \gamma^4 + 380 K_{fz}^3 \gamma^4 \nonumber\\&
+ 180 (315 + 511 K_{fz} + 810 \gamma)- 2 (2 K_{fz} (10 K_{fz}^2 (14 + 19 \gamma) (16 + \gamma (33 + \gamma (12 + \gamma))) \nonumber\\&
- 135 (-28 + \gamma (-42 + \gamma (48 + \gamma (88 + 21 \gamma)))) + 3 K_{fz} (2912 + \gamma (9138 + \gamma (8569 + \nonumber\\&
\gamma (2162 + 109 \gamma)))) ) + K_{fy}^2 (20 K_{fz} (14 + 19 \gamma) (16 + \gamma (33 + \gamma (12 + \gamma))) + 9 (896 + \nonumber\\&
\gamma (3244 + \gamma (3782 + \gamma (1556 + 217 \gamma)))) ) ) \tilde{\rho} + ( 4 K_{fz}^2 (252 + 5 K_{fz} (14 + 19 \gamma) (10 + \nonumber\\&
\gamma (19 + \gamma (8 + \gamma))) - 9 \gamma (12 + \gamma (219 + \gamma (250 + 61 \gamma))))  + K_{fy}^2 ( 20 K_{fz} (14 + 19 \gamma) (10 + \nonumber\\&
\gamma (19 + \gamma (8 + \gamma))) + 9 (644 + \gamma (2304 + \gamma (2803 + 7 \gamma (190 + 31 \gamma)))) ) ) \tilde{\rho}^2 + K_{fx}^2 \nonumber\\&
( 20 K_{fz} (14 + 19 \gamma) (18 + \gamma (45 + \gamma (16 + \gamma)) - 32 \tilde{\rho}  - 2 \gamma (33 + \gamma (12 + \gamma)) \tilde{\rho} + (10 +\nonumber\\&
\gamma (19 + \gamma (8 + \gamma))) \tilde{\rho}^2) + 9 ( 217 \gamma^4 (-1 + \tilde{\rho})^2 + 2 \gamma^3 (-1 + \tilde{\rho}) (-891 + 665 \tilde{\rho})  + 28 (33 + \nonumber\\&
\tilde{\rho} (-64 + 23 \tilde{\rho})) + 8 \gamma (471 + \tilde{\rho} (-811 + 288 \tilde{\rho})) + \gamma^2 (4609 + \tilde{\rho} (-7564 + 2803 \tilde{\rho})) ) ) ).\label{eq:F2un_z gravity arbitrary}
\end{align}
\section{Modification in the analysis of Murata  \citep{murata1980deformation}}\label{app:modification in Murata analysis}
Murata calculated the velocity of a weakly elastic sphere sedimenting under gravity in an unbounded domain. The velocity of the particle ($V_{p}\mathbf{\hat{k}}$) is in the direction of gravity ($g_z\mathbf{\hat{k}}$). In our notation, the velocity can be written as an asymptotic series in $\alpha$ as
\begin{align}
    V_{p} = \frac{2[\tilde{\rho}-1]\rho_fg_zR_0^2}{9 \mu}\left[1+\alpha^2 V^{(2)}\right].
\end{align}
Here, $V^{(2)}$ is the velocity coefficient at \textit{O}($\alpha^2$) and is given by
\begin{align}
     V^{(2)} =\frac{-27 \big(6 + \gamma [18 + \gamma]\big) + 6 K_{fz} (-1 + 16 \gamma) \tilde{\rho} + 8 K_{fz}^2 \tilde{\rho}^2}{20 (2 + 3 \gamma)^2}. \label{appeq:V2 in murata}
\end{align}
Murata had neglected the density variation of the particle. If one includes that, the coefficient of the terminal velocity at \textit{O}($\alpha^2$) is different from (\ref{appeq:V2 in murata}). The modified coefficient at \textit{O}($\alpha^2$) is obtained as
\begin{align}
   V^{(2)}_{mod} = \frac{-243 \big(6 + \gamma [18 + \gamma]\big) + 18 K_{fz} \big(32 + \gamma [133 + 5 \gamma]\big) \tilde{\rho} + 6 K_{fz}^2 (7 - 50 \gamma) \tilde{\rho}^2 - 20 K_{fz}^3 \tilde{\rho}^3}{180 (2 + 3 \gamma)^2}.
\end{align}

%
\end{document}